\documentclass[useAMS,usenatbib]{mn2e}

\usepackage{color}
\usepackage{colortbl}
\usepackage{multirow}
\usepackage{dcolumn}
\usepackage{amsmath}
\usepackage{amssymb}
\usepackage{graphicx}
\usepackage{epsfig}
\usepackage{changebar}
\usepackage{natbib}
\usepackage{multirow}
\usepackage{subfigure}
\usepackage{txfonts}
\usepackage{rotating}
\usepackage{threeparttable}
\usepackage{ulem}

\usepackage{longtable,lscape}

\newcolumntype{d}[1]{D{.}{\cdot}{#1}}

\newcolumntype{.}{D{.}{.}{-1}}

\newcommand{\lsun}{\ensuremath{\mathrm{L}_\odot}}
\newcommand{\msun}{M$_\odot$}

\newcommand{\vlsr}{$V_{\rm{LSR}}$}

\newcommand{\mum}{$\umu$m}

\newcommand{\kms}{km\,s$^{-1}$}

\newcommand{\hi}{H~{\sc i}}
\newcommand{\hii}{H~{\sc ii}}
\newcommand{\uchii}{UC\,H~{\sc ii}}

\newcommand{\poi}{Poisson}
\newcommand{\Tr}{$T^*_{\rm{R}}$}

\newcommand{\CO}{\ensuremath{^{12}}CO}
\newcommand{\tCO}{\ensuremath{^{13}}CO}
\newcommand{\CeO}{C\ensuremath{^{18}}O}

\newcommand{\NHthree}{NH$_3$}

\newcommand{\rms}{r.m.s.}

\newcommand{\Nor}{Norma}
\newcommand{\Scu}{Scutum-Centaurus}
\newcommand{\Sag}{Sagittarius}
\newcommand{\Per}{Perseus}

\newcommand{\zo}{$Z{\rm{o}}$}

\title[Structure of the Galaxy]{The RMS Survey: Galactic distribution of massive star formation\thanks{The full version of Tables\,1, 2 and 3 and Figs\,1 and 4 are only available in electronic form at the CDS via anonymous ftp to cdsarc.u-strasbg.fr (130.79.125.5) or via http://cdsweb.u-strasbg.fr/cgi-bin/qcat?J/MNRAS/.}}
\author[J. S. Urquhart et al.]{J.\,S.\,Urquhart$^{1}$\thanks{E-mail:
jurquhart@mpifr-bonn.mpg.de (MPIfR)}, C.\,C.\,Figura$^{2}$, T.\,J.\,T.\,Moore$^{3}$ M.\,G.\,Hoare${^4}$, S.\,L.\,Lumsden${^4}$, J.\,C.\,Mottram$^{5}$, \newauthor M.\,A.\,Thompson$^{6}$ \& R.\,D.\,Oudmaijer$^{4}$ \\
\\
$^{1}$
 Max-Planck-Institut f\"ur Radioastronomie, Auf dem H\"ugel
  69, D-53121 Bonn, Germany \\
$^{2}$Wartburg College, 100 Wartburg Blvd, Waverly, IA 50677, USA\\ 
$^{3}$Astrophysics Research Institute, Liverpool John Moores University, 146 Brownlow Hill, Liverpool, L3\,5RF, UK\\ 
$^{4}$School of Physics and Astrophysics, University of Leeds, Leeds, LS2\,9JT, UK \\
$^{5}$Leiden Observatory, Leiden University, NL-2300 RA Leiden, the Netherlands\\
 $^{6}$ Centre for Astrophysics Research, Science and Technology Research Institute, University of Hertfordshire, College Lane, Hatfield, AL10 9AB, UK \\ 
 }

\begin{document}

\date{Accepted ??. Received ??; in original form ??}

\pagerange{\pageref{firstpage}--\pageref{lastpage}} \pubyear{2009}

\maketitle

\label{firstpage}

\begin{abstract}

We have used the well-selected sample of $\sim$1750 embedded, young, massive stars identified by the Red MSX Source (RMS) survey to investigate the Galactic distribution of recent massive star formation. We present molecular-line observations for $\sim$800 sources without existing radial velocities. We describe the various methods used to assign distances extracted from the literature, and solve the distance ambiguities towards approximately 200 sources located within the Solar circle using archival \hi\ data. These distances are used to calculate bolometric luminosities and estimate the survey completeness ($\sim2\times10^4$\,\lsun). In total, we calculate the distance and luminosity of $\sim${\color{black}1650} sources, one third of which are above the survey's completeness threshold. Examination of the sample's longitude, latitude, radial velocities and mid-infrared images has identified $\sim$120 small groups of sources, many of which are associated with well known star formation complexes, such as G305, G333, W31, W43, W49 and W51.
	
	We compare the positional distribution of the sample with the expected locations of the spiral arms, assuming a model of the Galaxy consisting of four gaseous arms. The distribution of young massive stars in the Milky Way is spatially correlated with the spiral arms, with strong peaks in the source position and luminosity distributions at the arms' Galactocentric radii. The overall source and luminosity surface densities are both well correlated with the surface density of the molecular gas, which suggests that the massive star formation rate (SFR) per unit molecular mass is approximately constant across the Galaxy. A comparison of the distribution of molecular gas and the young massive stars to that in other nearby spiral galaxies shows similar radial dependencies. 
	
	We estimate the total luminosity of the embedded massive star population to be $\sim0.76\times10^8$\,\lsun, 30\,per\,cent of which is associated with the ten most active star forming complexes. We measure the scale height as a function of Galactocentric distance and find that it increases only modestly from $\sim$20-30\,pc between 4 and 8\,kpc, but much more rapidly at larger distances.

\end{abstract}

\section{Introduction}
\label{sect:intro}

Although significantly more luminous than their low-mass counterparts, massive stars ($M_\star>8$\,\msun) pose a challenging problem for study, particularly in their early formation stages.  Massive stars are much less common and have shorter lifetimes than intermediate- and low-mass stars. They are known to form almost exclusively in clusters \citep{de-wit2004} where source confusion limits the ability to discriminate between individual stars.  Because these stars form much more quickly than intermediate- and low-mass stars\citep[e.g.,][]{davies2011,mottram2011b}, they reach the main sequence while still deeply embedded within their natal clump, so that their formation stages take place beneath hundreds of magnitudes of visual extinction.  These observational difficulties complicate the identification of the large statistical samples required to investigate the earliest stages of massive star formation: as a result, our understanding of the initial conditions required or the processes involved in massive star formation is much poorer than for lower-mass stars.

The motivation for understanding the formation of massive stars is nevertheless strong: these stars are responsible for many of the higher-energy events in the universe, and play a significant role in the evolution of their host galaxies \citep{kennicutt2005}. Throughout their lives they enrich the local chemistry, and inject an enormous amount of radiative and mechanical energy into the interstellar medium (ISM) in the form of UV radiation, stellar winds, jets and outflows, and supernova explosions. 

These feedback mechanisms play a role in regulating the star formation in their vicinity by disrupting molecular clouds before star formation has begun, or by triggering the formation of future generations of stars in the surrounding molecular material.  Triggering may occur by sweeping up and  compressing the molecular material in surrounding clouds via the collect-and-collapse mechanism (e.g., \citealt{whitworth1994, deharveng2003}) or through radiatively-driven implosion (e.g., \citealt{bertoldi1989, urquhart2007d}). The processes involved in the formation of massive stars and their subsequent impact on their local environment is a key element  to understanding the role massive stars play in  shaping the dynamics and structure of their host galaxies. 

The Red MSX Source (RMS; \citealt{hoare2005,urquhart2008}; Lumsden et al. 2013, submitted) survey has been tailored to address many of these outstanding questions by identifying a large well-selected sample of massive young stellar objects (MYSOs) and compact and ultra-compact (UC) \hii\ regions. The RMS survey essentially consists of a suite of  follow-up observations and complementary data from other surveys (e.g., 2MASS, UKIDSS, VVV, GLIMPSE, MIPSGAL, ATLASGAL and CORNISH; \citealt{2mass}, \citealt{vvv}, \citealt{lawrence2007}, \citealt{benjamin2003_ori}, \citealt{carey2009}, \citealt{schuller2009} and \citealt{hoare2012}, respectively) of a mid-infrared colour selected sample of $\sim$5000 MSX sources \citep{lumsden2002}. This initial sample contains a significant number of dusty objects such as evolved stars that have similar mid-infrared colours as embedded young stars; however, the follow-up observations had been carefully chosen to identify and remove these contaminating sources from the final sample. A database has been constructed to hold all of these multi-wavelength data sets and to compile all of the available data on a source-by-source basis to aid in their classification.\footnote{\tt{http://rms.leeds.ac.uk/cgi-bin/public/RMS\_DATABASE.cgi.}} 

With our programme of follow-up observations and the classification now complete, we have identified approximately 1600 YSO and \hii\ regions located throughout the Galactic plane ($|{b}| < 5\degr$). This sample of young embedded massive stars is an order of magnitude larger than was previously available. The RMS sample provides a sufficient number of sources to allow statistically significant studies of young massive stars as a function of luminosity and environment, while avoiding many of the biases associated with previous surveys of young massive stars. A complete overview of this project, detailed description of the classification scheme, and discussion of the properties of the final embedded catalogue is presented in Lumsden et al. (2013). 

We have investigated the Galactic distribution of RMS MYSOs in two previous papers, both of which were focused on smaller subsamples \citep[i.e.,][]{urquhart2011a,urquhart2012}. In this paper we build on these previous studies and use the full RMS sample to provide a comprehensive picture of the Galactic distribution of massive star formation. We will compare this distribution of the now complete RMS sample of massive young stars with the expected positions of the spiral arms, and investigate the source and luminosity surface densities as a function of Galactic location. 

In the next section we will summarise the previous molecular line observations and describe additional observations that have been undertaken in an effort to obtain a complete set of radial velocities for this sample. In Section\,\ref{sect:distances} we review the various methods used to determine distances, which are then used to calculate individual source bolometric luminosities and estimate the survey's completeness. We use these distances and luminosities to investigate the Galactic distribution of massive star formation and to compare this with the position of the spiral arms and measure the scaleheight of the Galactic disk in Section\,\ref{sect:gal_structure}. Finally, in Section\,\ref{sect:summary_conclusions} we present a summary of our results and highlight our main findings.

\section{Radial velocity measurements}
\subsection{Previous observations}

Molecular-line observations are a crucial part of the RMS follow-up campaign, as they provide radial velocities for the determination of kinematic distances and luminosities. These are used to distinguish nearby low- and intermediate-mass YSOs from the generally more distant MYSOs. They can also be useful in identifying the more evolved stars that contaminate our sample, as these are not generally associated with sufficient  cold gas to produce a detectable emission in the lower excitation states \citep[i.e.,][]{loup1993}.

The \tCO\ (1-0) and (2-1) transitions were chosen as optimal tracers as they have a lower abundance than their $^{12}$CO counterparts.  \CO\ is generally optically thick in these regions, and suffers from saturation, self-absorption and multiple emission components along a given line-of-sight that can produce complex line profiles, while \tCO\ is generally free from these problems.  Furthermore, \tCO\ is more abundant than \CeO, allowing emission to be detected in a modest amount of observing time. We have made \tCO\ observations towards $\sim$2000 sources; the results of the majority of these were reported in \citet{urquhart_13co_south,urquhart_13co_north} and the remainder will be discussed in the following subsection. 

Although the $^{13}$CO transition is much less affected by many of the problems associated with $^{12}$CO emission, multiple components are still detected towards $\sim$60\,per\,cent of the observed sources. We have therefore followed up a large number of these with observations of high density tracers such as CS~(2-1), NH$_3$ and water maser observations (\citealt{urquhart2009_h2o,urquhart2011b}). We have complemented these targeted observations with other high density transitions reported in the literature \citep[e.g.,][]{bronfman1996,schlingman2011,wienen2012}.

\subsection{Additional observations: $^{13}$CO, CS and NH$_3$}

\setlength{\tabcolsep}{6pt}

\begin{table*}

\begin{center}
\caption{ Fitted molecular line parameters.}
\label{tbl:addition_obs}
\begin{minipage}{\linewidth}
\begin{tabular}{lccc....}
\hline
\hline

\multirow{2}{24mm}{Field Name}	&	RA  		&Dec.  	& \multirow{2}{14mm}{Transition}	& \multicolumn{1}{c}{\rms} &	\multicolumn{1}{c}{\vlsr} 	& \multicolumn{1}{c}{\Tr}	& \multicolumn{1}{c}{FWHM}\\ 
&	(J2000) 	& (J2000) 	& & \multicolumn{1}{c}{(mK)}  		& \multicolumn{1}{c}{\kms} &\multicolumn{1}{c}{(K)} 	&\multicolumn{1}{c}{(\kms)} 	   \\  

\hline

G305.2017+00.2072	&	13:11:10.29	&	$-$62:34:39.0	&	NH$_{3}$ (1,1)	&	15	&	-41.38	&	0.43	&	12.24	\\
G305.2017+00.2072	&	13:11:10.29	&	$-$62:34:39.0	&	NH$_{3}$ (2,2)	&	16	&	-41.54	&	0.18	&	5.94	\\
G305.2242+00.2028	&	13:11:22.07	&	$-$62:34:48.0	&	CS (J=2-1)	&	60	&	-40.69	&	0.31	&	9.65	\\
G305.2242+00.2028	&	13:11:22.12	&	$-$62:34:48.7	&	NH$_{3}$ (1,1)	&	16	&	-41.45	&	0.09	&	10.29	\\
G305.2242+00.2028	&	13:11:22.12	&	$-$62:34:48.7	&	NH$_{3}$ (2,2)	&	14	&	-41.81	&	0.08	&	5.84	\\
G305.2535+00.2412	&	13:11:35.80	&	$-$62:32:22.9	&	NH$_{3}$ (1,1)	&	14	&	-36.91	&	0.04	&	10.84	\\
G305.2535+00.2412	&	13:11:35.80	&	$-$62:32:22.9	&	NH$_{3}$ (2,2)	&	13	&	-38.55	&	0.08	&	3.66	\\
G305.3210+00.0706	&	13:12:17.96	&	$-$62:42:16.3	&	$^{13}$CO (J=1-0)	&	71	&	-31.71	&	0.33	&	2.83	\\
G305.3210+00.0706	&	13:12:17.96	&	$-$62:42:16.3	&	$^{13}$CO (J=1-0)	&	71	&	-38.80	&	0.74	&	5.19	\\
G305.3210+00.0706	&	13:12:17.96	&	$-$62:42:16.3	&	$^{13}$CO (J=1-0)	&	71	&	-45.32	&	0.59	&	2.40	\\
G305.3526+00.1945	&	13:12:29.36	&	$-$62:34:41.2	&	$^{13}$CO (J=1-0)	&	67	&	-37.69	&	6.20	&	6.70	\\
G305.3676+00.2095	&	13:12:36.44	&	$-$62:33:43.5	&	$^{13}$CO (J=1-0)	&	65	&	-34.64	&	6.58	&	4.82	\\
G305.3719+00.1837	&	13:12:39.69	&	$-$62:35:15.0	&	NH$_{3}$ (1,1)	&	16	&	-38.11	&	0.25	&	9.03	\\
G305.3719+00.1837	&	13:12:39.69	&	$-$62:35:15.0	&	NH$_{3}$ (2,2)	&	15	&	-38.48	&	0.13	&	5.48	\\
G305.3779+00.2108	&	13:12:41.56	&	$-$62:33:35.4	&	$^{13}$CO (J=1-0)	&	67	&	-33.57	&	5.16	&	3.46	\\
G305.3779+00.2108	&	13:12:41.56	&	$-$62:33:35.4	&	$^{13}$CO (J=1-0)	&	67	&	-38.53	&	1.37	&	5.59	\\
G305.3779+00.2108	&	13:12:41.56	&	$-$62:33:35.4	&	$^{13}$CO (J=1-0)	&	67	&	-45.27	&	0.87	&	0.98	\\
G305.4399+00.2103	&	13:13:13.84	&	$-$62:33:19.0	&	NH$_{3}$ (1,1)	&	15	&	\multicolumn{1}{c}{$\cdots$}	&	\multicolumn{1}{c}{$\cdots$}	&	\multicolumn{1}{c}{$\cdots$}	\\
G305.4399+00.2103	&	13:13:13.84	&	$-$62:33:19.0	&	NH$_{3}$ (2,2)	&	15	&	\multicolumn{1}{c}{$\cdots$}	&	\multicolumn{1}{c}{$\cdots$}	&	\multicolumn{1}{c}{$\cdots$}	\\
G305.4748$-$00.0961	&	13:13:45.76	&	$-$62:51:27.7	&	NH$_{3}$ (1,1)	&	10	&	-38.36	&	0.25	&	5.17	\\
G305.4748$-$00.0961	&	13:13:45.76	&	$-$62:51:27.7	&	NH$_{3}$ (2,2)	&	11	&	-38.48	&	0.09	&	2.35	\\
G305.4840+00.2248	&	13:13:35.99	&	$-$62:32:12.8	&	CS (J=2-1)	&	46	&	-44.59	&	0.55	&	1.41	\\
G305.4840+00.2248	&	13:13:36.05	&	$-$62:32:13.5	&	NH$_{3}$ (1,1)	&	10	&	-44.77	&	0.07	&	1.92	\\
G305.4840+00.2248	&	13:13:36.05	&	$-$62:32:13.5	&	NH$_{3}$ (2,2)	&	11	&	\multicolumn{1}{c}{$\cdots$}	&	\multicolumn{1}{c}{$\cdots$}	&	\multicolumn{1}{c}{$\cdots$}	\\
G305.5393+00.3394	&	13:13:59.52	&	$-$62:25:05.5	&	NH$_{3}$ (1,1)	&	13	&	-35.13	&	0.70	&	8.40	\\
G305.5393+00.3394	&	13:13:59.52	&	$-$62:25:05.5	&	NH$_{3}$ (2,2)	&	13	&	-35.17	&	0.05	&	2.94	\\
G305.5516+00.0149	&	13:14:20.94	&	$-$62:44:26.4	&	$^{13}$CO (J=1-0)	&	90	&	-31.06	&	1.46	&	1.91	\\
G305.5516+00.0149	&	13:14:20.94	&	$-$62:44:26.4	&	$^{13}$CO (J=1-0)	&	90	&	-38.88	&	6.30	&	3.46	\\
G305.5610+00.0124	&	13:14:25.82	&	$-$62:44:30.8	&	NH$_{3}$ (1,1)	&	14	&	-39.13	&	0.09	&	3.63	\\
G305.5610+00.0124	&	13:14:25.82	&	$-$62:44:30.8	&	NH$_{3}$ (2,2)	&	15	&	-38.42	&	0.06	&	3.11	\\
\hline
\end{tabular}

\end{minipage}

\end{center}
Notes: A small portion of the data is provided here, the full table is only  available in electronic form at the CDS via anonymous ftp to cdsarc.u-strasbg.fr (130.79.125.5) or via http://cdsweb.u-strasbg.fr/cgi-bin/qcat?J/MNRAS/.

\end{table*}
\setlength{\tabcolsep}{6pt}

The observations described in this section were made with the 22-m Mopra radio telescope, which is located near Coonabarabran, New South Wales, Australia.\footnote{The Mopra radio telescope is part of the Australia Telescope National Facility which is funded by the Commonwealth of Australia for operation as a National Facility managed by CSIRO.} The Mopra beam size is approximately 2.5\arcmin\ for the ammonia observations and 36\arcsec\ and 40\arcsec\ for the CO and CS observations, respectively.

The CS\,(2-1) observations were made with Mopra's older cryogenically-cooled ($\sim$4\,K) Superconductor-Insulator-Superconductor (SIS) junction mixer, with a frequency range between 85-116\,GHz.  The receiver backend was a digital autocorrelator capable of providing two simultaneous outputs with an instantaneous bandwidth between 4-256\,MHz.  The CS observations were made using a bandwidth of 64\,MHz with a 1024-channel digital autocorrelator. This provided a frequency resolution of 62.5\,kHz and a velocity resolution of $\sim$0.2\,\kms\ at the CS transitions rest frequency (97.98\,GHz).

The UNSW Mopra spectrometer (MOPS)\footnote{The University of New South Wales Digital Filter Bank used for the observations with the Mopra Telescope was provided with support from the Australian Research Council.} was commissioned in October 2005 and consists of four 2.2\,GHz bands that overlap slightly to provide a total of 8\,GHz continuous bandwidth. Up to four zoom windows can be placed within each 2.2\,GHz band allowing up to 16 spectral windows to be observed simultaneously, each providing a bandwidth of 137\,MHz with 4096 channels. MOPS was used for the  NH$_3$ (1,1) \& (2,2) inversion transitions,  and the \tCO\ (1-0) observations described in the following two subsections.

\subsubsection{CS and NH$_3$ observations}

\NHthree\ and CS observations were made towards sources previously observed in the $^{13}$CO transition as part of our initial programme of follow-up observations. As previously mentioned, multiple velocity components were seen along the lines of sight towards many of these sources, and further observations of higher-density tracers were required to identify the velocity component asoociated with the IR continuum source. The Mopra telescope was used to follow up a large number of sources using CS (2-1) and the two lower-excitation inversion transitions of NH$_3$. All of these transitions have high critical densities ($\sim$10$^4$--$10^5$\,cm$^{-3}$) and are therefore excellent tracers of high density molecular gas associated with high-mass star-forming regions. 

The CS observations were made towards {\color{black}127} sources in September 2004 and August 2005 (project reference: M121). The signal-to-noise ratio was improved by tuning both polarisations to the CS frequency. System temperatures were typically $\sim$200\,K.

The NH$_3$ observations were made in April and September 2008 towards {\color{black}499} RMS sources using the 12-mm receiver and MOPS (project reference: M270; \citealt{m270}). The observations covered a frequency range of 16 to 27.5\,GHz. Each of the 16 zoom windows was centred on the rest frequency of a known transition providing a total velocity coverage of $\sim$2000\,\kms\ with a resolution of $\sim$0.4\,\kms\ per channel. System temperatures were between 65 and 100\,K. A detailed description of these observations can be found in \citet{urquhart2009_h2o}. Although they included a large number of transitions, we only present the NH$_3$ (1,1) and (2,2) transitions here; however, we have made the data for all of the other transitions available in the RMS database.

\subsubsection{CO observations}

The RMS source classification criteria have evolved over time as the data obtained from the various programmes of follow-up observations have been analysed (we refer the reader to \citealt{lumsden2013} for a detailed description of the classification criteria is presented). This has resulted in a number of sources that were initially rejected being reclassified as either a YSO or \hii\ region. As these reintroduced sources were not included in our initial sample, no molecular-line observation had been previously made. 

We made observations of the CO (1-0) transitions towards 192 of the re-introduced RMS sources in March 2012 using the MOPS instrument (project reference: M573; \citealt{m573}). Three zoom windows were used to cover the $^{12}$CO, $^{13}$CO and C$^{18}$O transitions. However, the optically thick $^{12}$CO emission generally shows complex emission profiles while the C$^{18}$O is generally only detected towards a small number of sources. We therefore only present the plots of the $^{13}$CO emission profiles and the fit parameters to the emission features seen in these spectra. The average system temperature was $\sim$360\,K at 110\,GHz.

\begin{figure}
\begin{center}

\includegraphics[width=0.49\textwidth, trim= 0 0 0 0]{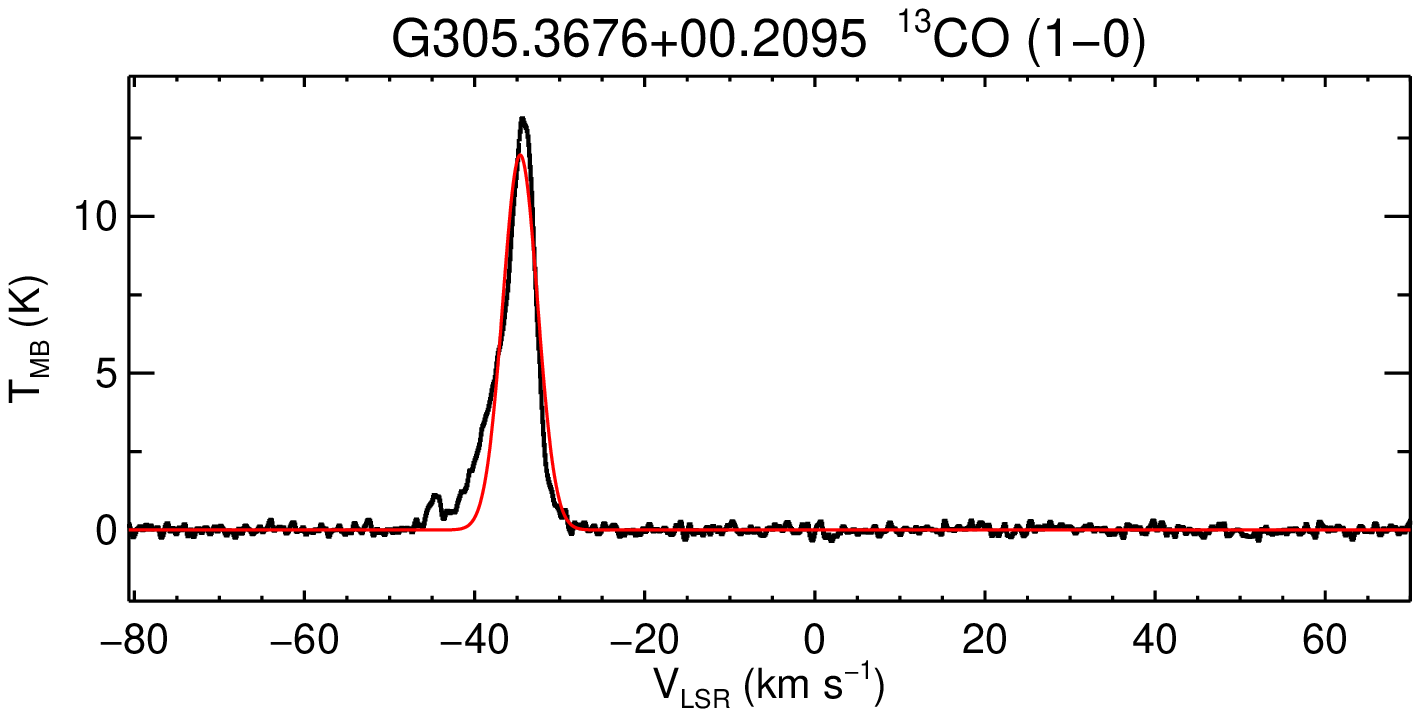}
\includegraphics[width=0.49\textwidth, trim= 0 0 0 0]{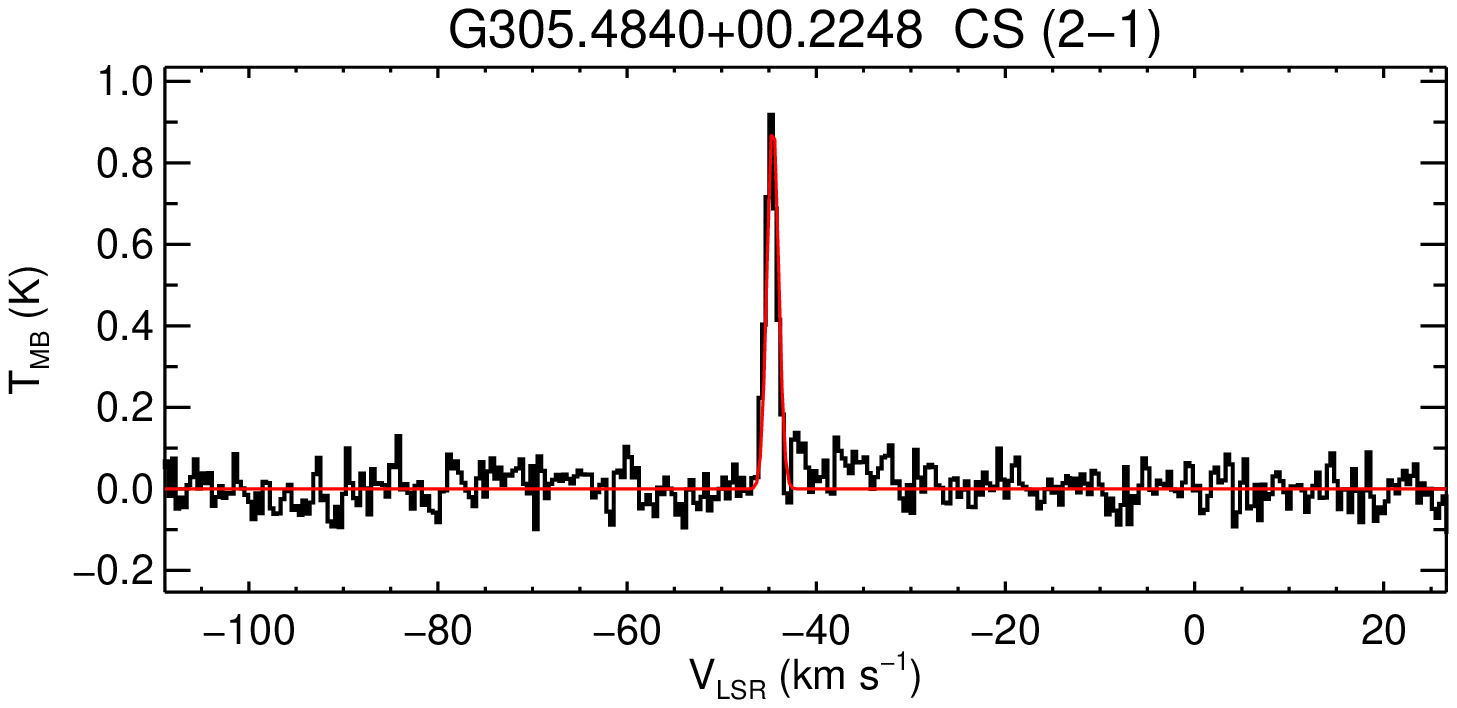}
\includegraphics[width=0.49\textwidth, trim= 0 0 0 0]{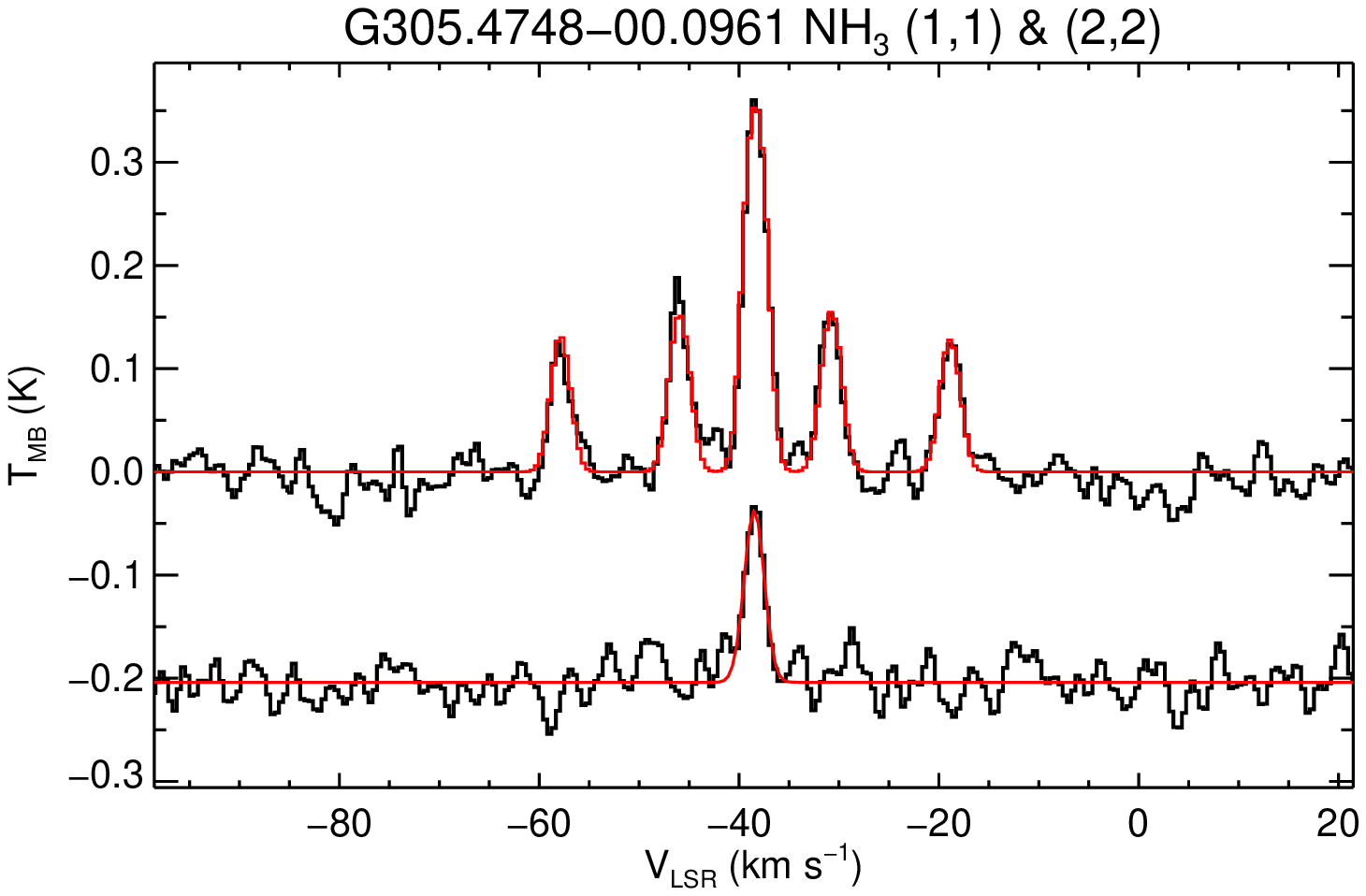}

\caption{\label{fig:example_spectra} Example spectra obtained from the additional Mopra observations. In the upper and middle panels we present emission in the CS and $^{13}$CO (1-0) transitions towards example RMS sources. The lower panel shows example NH$_3$ (1,1) and (2,2) inversion transitions, with the latter offset by $-$0.2\,K to avoid confusion. The hyperfine structure can be clearly in the NH$_3$ (1,1) emission profile. In all of these plots the black and red lines show the observed spectra and model fits to the data, respectively. All spectra have a velocity resolution of $\sim$0.4\,\kms.} 

\end{center}
\end{figure}

\subsection{Observational procedures and data reduction}

All molecular-line observations were performed in position-switching mode, with typical on-source integration times of $\sim$10, 30 and 40 minutes for the $^{13}$CO, NH$_3$ and CS transitions, respectively. The on-source integration time was split into individual one-minute on- and off-source scans. Reference positions were offset from source positions by 1 degree in a direction perpendicular to the Galactic plane. These were chosen to avoid contamination by emission at the reference position at a similar velocity.

Weather conditions were stable over the short time periods required to complete the observations of each source, with system temperatures varying by no more than approximately 10\,per\,cent. The telescope pointing was checked every 1-2 hours by observing a strong nearby SiO maser (\citealt{indermuehle2013}) and the average pointing accuracy was found to be better than 10\arcsec\ \rms. A measurement of an ambient load (assumed to be at 290\,K) was made before and after each $^{13}$CO and CS observation following the method of \citet{kutner1981} to put the measured antenna temperatures onto the standard $T^*_{\rm{A}}$ scale, correcting for atmospheric absorption, ohmic losses and rearward spillover and scattering. This correction is not required for observations below 70\,GHz as the atmosphere is significantly less variable and absorption from water vapour is less of a problem.

The observations of all three transitions were reduced using the ATNF spectral line reducing software (ASAP).  Individual on-off scans were processed to remove sky emission and visually inspected to remove poor scans.  A low-order polynomial baseline was fitted, after which the individual scans were averaged together to produce a single spectrum for each source. 

The CO and CS spectra have been Hanning smoothed to a velocity resolution of $\sim$0.4\,\kms\ to obtain a final sensitivity of $\sim$70 and 35\,mK\,channel$^{-1}$\,beam$^{-1}$, respectively. The reduced spectra were converted to the telescope-independent main-beam temperature scale ($T_{\rm{mb}}$), assuming a main-beam efficiency ($\eta_{\rm{mb}}$) of 0.65 \citep{urquhart2010a}, 0.60 and 0.55 \citep{ladd2005} for the NH$_3$, CS and CO observations, respectively. In Fig.\,\ref{fig:example_spectra} we present an example of each of the emission detected for the three molecules observed.

Gaussian profiles were fitted to the observed emission features in the CO and CS data using the spectral line analysis package \textsc{XS} written by Per Bergman.\footnote{Available from the Onsala Space Observatory at http://www.chalmers.se/rss/oso-en/observations/data-reduction-software.} Where necessary, a higher-order polynomial was fitted to the emission-free parts of the spectrum and subtracted from the baseline before the Gaussian profiles were fitted. As can be seen in the upper and middle panels of Fig.\,\ref{fig:example_spectra}, the Gaussian profile provided a good fit to the data in the majority of cases.

\setlength{\tabcolsep}{6pt}

\begin{table*}

\begin{center}
\caption{Parameters of the 25 most luminous star-forming complexes. The positions and velocities of these complexes have been determined from the mean of all associated RMS sources, and as such are only approximate values. The luminosities of each complex (given in Col.\,9) have been determined from the integrated bolometric luminosities of their embedded YSO and \hii\ region populations. The contribution each complex makes to the total embedded massive star population (Col.\,10) is estimated by dividing the complex's luminosity by the estimate of total Galactic MYSO and embedded \hii\ region luminosity ($L_{\rm{Galaxy}}=0.76\times 10^8$\,\lsun; see Sect.\,4.1 for details).}
\label{tbl:complex_parameters}
\begin{minipage}{\linewidth}
\begin{tabular}{l.........}
\hline
\hline

\multirow{2}{24mm}{Complex Name}	&	\multicolumn{1}{c}{$\ell$}  		&\multicolumn{1}{c}{$b$}  	& \multicolumn{1}{c}{\vlsr}	& \multicolumn{1}{c}{RMS Members} &	\multicolumn{1}{c}{Distance} 	& \multicolumn{1}{c}{z}	& \multicolumn{1}{c}{$R_{\rm{GC}}$}&\multicolumn{1}{c}{$L_{\rm{bol}}$} &\multicolumn{1}{c}{$L_{\rm{bol}}$/$L_{\rm{Galaxy}}$} \\
&	\multicolumn{1}{c}{(\degr)} 	& \multicolumn{1}{c}{(\degr)} 	&\multicolumn{1}{c}{(\kms)} & \multicolumn{1}{c}{(\#)}  		& \multicolumn{1}{c}{(kpc)} &\multicolumn{1}{c}{(pc)} 	&\multicolumn{1}{c}{(kpc)} 	& \multicolumn{1}{c}{(Log[\lsun])}& \multicolumn{1}{c}{(\%)}   \\  

\hline

W51\,(A\&B)	&	49.362	&	-0.330	&	60.47	&	24	&	5.4	&	-31.1	&	6.45	&	6.67	&	6.82	\\
NGC3603	&	291.598	&	-0.491	&	12.79	&	4	&	7.0	&	-59.6	&	8.78	&	6.61	&	5.95	\\
W49A	&	43.145	&	-0.009	&	11.14	&	10	&	11.1	&	-1.7	&	7.61	&	6.56	&	5.23	\\
RCW\,106 (G333)	&	333.037	&	-0.320	&	-51.65	&	33	&	3.6	&	-20.1	&	5.54	&	6.28	&	2.74	\\
G338.398+00.164	&	338.404	&	0.120	&	-35.74	&	14	&	12.8	&	28.9	&	5.82	&	6.21	&	2.36	\\
GAL331.03$-$00.15	&	330.960	&	-0.185	&	-91.54	&	2	&	9.6	&	-30.9	&	4.64	&	6.04	&	1.57	\\
G305	&	305.506	&	0.085	&	-36.32	&	25	&	4.0	&	5.8	&	6.98	&	5.94	&	1.26	\\
G282.0$-$1.2	&	281.881	&	-1.605	&	-6.10	&	11	&	7.0	&	-195.4	&	9.82	&	5.93	&	1.24	\\
W43	&	30.861	&	-0.023	&	93.73	&	30	&	5.1	&	-1.8	&	4.99	&	5.93	&	1.23	\\
AGAL032.797+00.191	&	32.796	&	0.198	&	14.99	&	2	&	12.9	&	44.5	&	7.37	&	5.89	&	1.12	\\
W47	&	37.601	&	-0.223	&	53.48	&	6	&	9.9	&	-38.3	&	6.07	&	5.88	&	1.10	\\
RCW42 	&	273.982	&	-1.191	&	37.52	&	5	&	5.5	&	-114.6	&	9.81	&	5.86	&	1.05	\\
AGAL045.121+00.131	&	45.115	&	0.131	&	58.83	&	2	&	4.4	&	10.1	&	6.23	&	5.82	&	0.96	\\
W3	&	133.797	&	1.155	&	-43.42	&	9	&	2.0	&	39.3	&	9.95	&	5.80	&	0.92	\\
AGAL319.399$-$00.012	&	319.390	&	-0.007	&	-13.50	&	4	&	11.6	&	-1.5	&	7.57	&	5.69	&	0.70	\\
Far 3$-$kpc arm (south)	&	348.854	&	-0.015	&	13.30	&	8	&	11.3	&	-2.8	&	3.36	&	5.67	&	0.67	\\
Near 3$-$kpc tangent	&	336.880	&	0.011	&	-123.24	&	10	&	7.7	&	1.6	&	3.34	&	5.56	&	0.53	\\
AGAL020.081$-$00.136	&	20.076	&	-0.139	&	41.80	&	2	&	12.6	&	-30.5	&	5.44	&	5.55	&	0.51	\\
GAL331.5$-$00.1	&	331.553	&	-0.088	&	-86.67	&	9	&	4.9	&	-7.7	&	4.81	&	5.52	&	0.48	\\
RCW116B	&	345.091	&	1.543	&	-14.77	&	14	&	2.4	&	64.1	&	6.23	&	5.43	&	0.39	\\
NGC7538	&	111.626	&	0.761	&	-55.48	&	9	&	2.6	&	35.2	&	9.79	&	5.43	&	0.38	\\
Cygnus-X	&	79.538	&	0.942	&	1.66	&	74	&	1.4	&	23.0	&	8.36	&	5.40	&	0.36	\\
Gum\,50	&	328.567	&	-0.533	&	-46.57	&	1	&	3.0	&	-28.3	&	6.12	&	5.38	&	0.35	\\
G010.960+00.017	&	10.962	&	0.015	&	20.29	&	2	&	8.1	&	2.8	&	5.76	&	5.36	&	0.33	\\
GAL336.40$-$00.23	&	336.470	&	-0.232	&	-86.11	&	5	&	10.5	&	-42.4	&	4.34	&	5.30	&	0.29	\\

\hline
\end{tabular}

\end{minipage}
\end{center}
Notes: A small portion of the data is provided here, the full table is only  available in electronic form at the CDS via anonymous ftp to cdsarc.u-strasbg.fr (130.79.125.5) or via http://cdsweb.u-strasbg.fr/cgi-bin/qcat?J/MNRAS/.

\end{table*}
\setlength{\tabcolsep}{6pt}

The NH$_3$ (1,1) molecule has hyperfine structure and consists of 18 components which need to be fitted simultaneously in order to derive the optical depth and line width (see lower panel of Fig.\,\ref{fig:example_spectra} for an example). The fitting of these data has been done in the IDL environment using the MPFIT routine.\footnote{MPFIT is part of a suite of routines written by Craig B. Markwardt and made available to the general public (http://www.physics.wisc.edu/$\sim$craigm/idl/idl.html).} The hyperfine structure is generally too weak to be observed in the NH$_3$ (2,2) transition, and so these detections have been fitted with Gaussian profiles along with the weaker NH$_3$ (1,1) lines where the hyperfine components are not detected. In all cases the NH$_3$ (1,1) and (2,2) line widths have been obtained by fitting the hyperfine components to their respective main line emission to remove the effects of line broadening due to optical depth effects.

The radial velocities, main beam temperatures and full-width at half-maximum (FWHM) line widths obtained from these fits are presented for all detected transitions in Table\,\ref{tbl:addition_obs}.

\begin{figure}
\begin{center}

\includegraphics[width=0.45\textwidth, trim= 0 0 0 0]{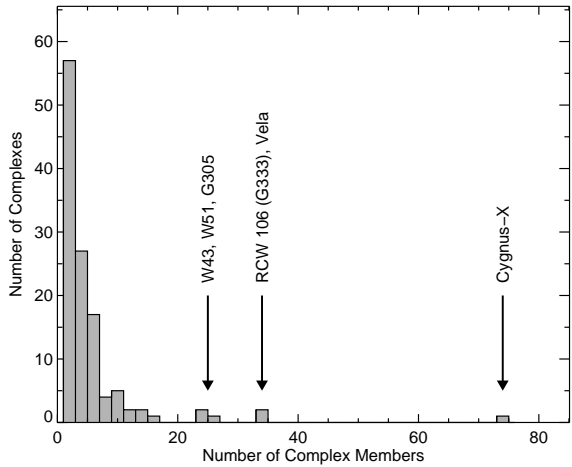}

\caption{\label{fig:cluster_member_distribution} Number of sources associated with each complex. The names of the complexes with the largest number of RMS associations are given.} 

\end{center}
\end{figure}

\subsection{Association of sources with known star-forming regions}
\label{sect:complexes}

Examination of $Spitzer$ mid-infrared images obtained as part of the GLIMPSE legacy project reveals that many RMS sources are associated with large star-forming complexes. In many cases, a single RMS source is associated with a previously identified complex but, in the majority of cases, we find multiple sources positionally coincident with a given complex. To confirm such associations, we compare their radial velocities and require that these agree within $\sim$10\,\kms. Since many complexes are well known (such as W31, M16, M17, etc) and have been the focus of detailed studies, reliable distances can be readily found in the literature (although care must be exercised regarding the comparison of different distance determinations, for example Westerlund\,2; \cite{dame2007} and W51\citet{clark2009} --- we have only adopted these in cases where these are broadly consistent).

\begin{figure*}
\begin{center}

\includegraphics[angle=90,width=0.32\textwidth, trim= 30 0 0 10]{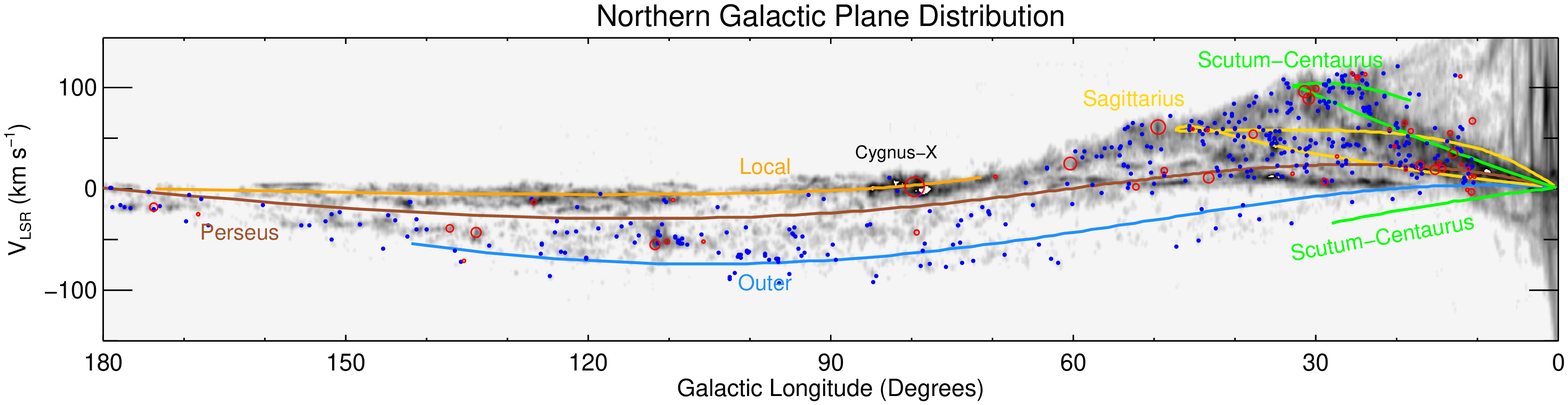}
\includegraphics[angle=90,width=0.32\textwidth, trim= 30 0 0 10]{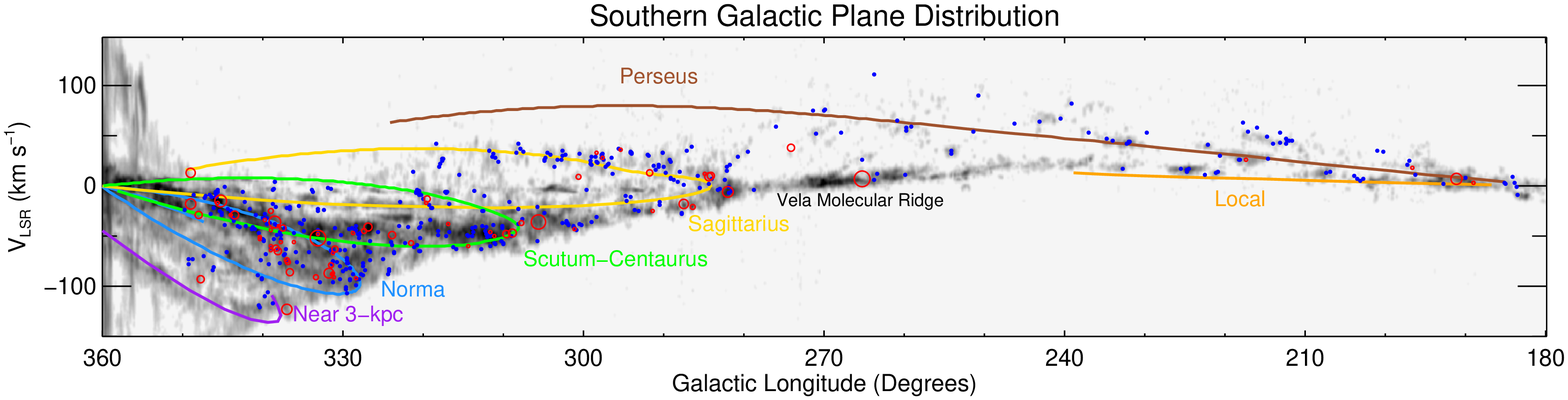}

\caption{\label{fig:lv_distribution} Left and right panels show the longitude and velocity distribution of the RMS sources in the Northern and Southern Galactic plane, respectively. The complexes are shown as open red circles, the size of which give an indication of the source density of each complex, while the blue filled circles show the position of the rest of the sample. The greyscale image shows the distribution of molecular gas as traced by the integrated $^{12}$CO emission (\citealt{dame2001}) for comparison. The location of the spiral and local arms are shown as curved solid lines, coloured to identify the individual arms. The positions of the four main spiral and local arms have been taken from model by \citet{tayor1993} and updated by \citet{cordes2004}, while the position of the near 3-kpc arm has been taken from \citet{bronfman2000}.}

\end{center}
\end{figure*}






We have used visual inspection of the mid-infrared images combined with velocity information to identify all of these small groups of sources and, where possible, have associated these groups with a known star-forming complex. This has resulted in $\sim$600 sources being associated with approximately 120 known star formation regions/complexes and accounts for $\sim$40\,per\,cent of the RMS sample of YSOs and \hii\ regions.  

Fig.\,\ref{fig:cluster_member_distribution} shows the number of RMS sources associated with each of the 117 complex identified and Table\,\ref{tbl:complex_parameters} lists the complex parameters. The vast majority of these complexes are relatively small, consisting of no more than a handful of individual massive star-forming regions. However, given that each one of these regions is probably forming a stellar cluster, these complexes should be viewed as active star forming regions. If it is typical for a given star-forming complex to have a small number of localised regions of active star formation, then complexes with significantly larger numbers of RMS sources stand out as perhaps being outside the norm. 

There are only six complexes that have more than 20 members: these are W51, Cygnus-X, W43, G305, RCW\,106 and the Vela molecular ridge, which are among the best-studied star-formation regions in the Galaxy (e.g., \citealt{parsons2012}, \citealt{csengeri2011}, \citealt{motte2003}, \citealt{hindson2010}, \citealt{bains2006} and \citealt{hill2012}, respectively). We note, however, that although Cygnus-X and the Vela molecular ridge are associated with the largest number of RMS sources, these are both relatively nearby (1.4 and 0.7\,kpc, respectively) and so the RMS sources are not particularly luminous.  Consequently, neither of these star forming complexes features in our list of the most luminous complexes presented in Table\,\ref{tbl:complex_parameters}. There is therefore a distance bias associated with the number of sources associated with each complex and care needs to be taken when drawing conclusions from the source counts alone.

By grouping sources together and identifying their associated star forming complexes, we have reduced the number of distances we need to find by $\sim$500; however, this still leaves some 1100 distances to be determined.  The methods used to determine these distances are discussed in Sect.\,\ref{sect:distances}.

\subsection{Longitude-velocity distribution}

Fig.\,\ref{fig:lv_distribution} shows the distribution of the complexes and individual embedded sources as a function of Galactic longitude and radial velocity. The upper and lower panels show the longitudinal distribution of the radial velocity (plotted over the integrated $^{12}$CO (1-0) emission mapped by \citealt{dame2001}) for the northern and southern Galactic plane, respectively. These longitude-velocity ($\ell$-$v$) plots also show the proposed positions of the four-arm spiral arm model of \citet{tayor1993} (updated by \citealt{cordes2004}), which itself is based on the earlier model of \citet{georgelin1976}.
The figure reveals a strong correlation between the distributions of molecular gas and RMS sources within the inner Galaxy (i.e., $|\ell |<60$\degr). Outside of this region, the correlation is significantly weaker.  The positions of the modelled spiral arms appears to be strongly correlated to the distribution of our sample of YSOs and \hii\ regions, particularly in the outer Galaxy where the correlation with the molecular gas is significantly poorer. Since the spiral arms in the outer Galaxy are the most distant, the poorer correlation between the RMS sources and the molecular gas there is likely to be due to the sensitivity of the CO survey, with the emission from the more distant molecular clouds being diluted in the 8\arcmin\ beam used by the Dame et al. survey (cf. Urquhart et al. 2013b). 

Previous CO, \hii\ region, and mid-infrared surveys have identified the spiral-arm line-of-sight tangents (e.g., \citealt{grabelsky1987,bronfman1988, alvarez1990,caswell1987,benjamin2008}). In the northern Galactic plane, tangent positions have been reported at $\ell\simeq$ 30\degr\ and 47\degr, which correspond to the \Scu\ and \Sag\ spiral arms. Inspection of the distribution of RMS sources in the northern Galactic plane reveals high densities towards the longitudes and velocities of these two tangents and their associations with W43 and W51 star forming complexes. Similarly, we find high RMS source densities towards the longitudes and velocities of the \Sag, \Scu, \Nor\, and near 3-kpc arms ($\sim$283\degr, 308\degr, 328\degr\ and 337\degr; \citealt{bronfman2000}).  We find G305 and G282.0$-$0.12 star forming complexes are associated with the \Scu\ and \Sag\ tangents, respectively. 

The source counts are significantly lower and broadly flat outside the spiral-arm tangents (i.e., $60\degr<\ell < 280\degr$) with only two notable exceptions; these are the Cygnus-X region ($\ell \sim$80\degr; \citealt{reipurth2008}) and the Vela molecular ridge ($\ell \sim$268\degr; \citealt{netterfield2009}). Both of these regions are likely to be associated with the local arm and are therefore relatively nearby (these two regions are labelled in the plots presented in Fig.\,\ref{fig:lv_distribution}). Incidentally, these two regions are also associated with two highest densities of RMS sources; however, as mentioned in the previous section, most of these are not very luminous. 

\section{Distances and luminosities}
\label{sect:distances}

\subsection{Distance determination}

\begin{figure}
\begin{center}

\includegraphics[width=0.49\textwidth, trim= 0 0 0 0]{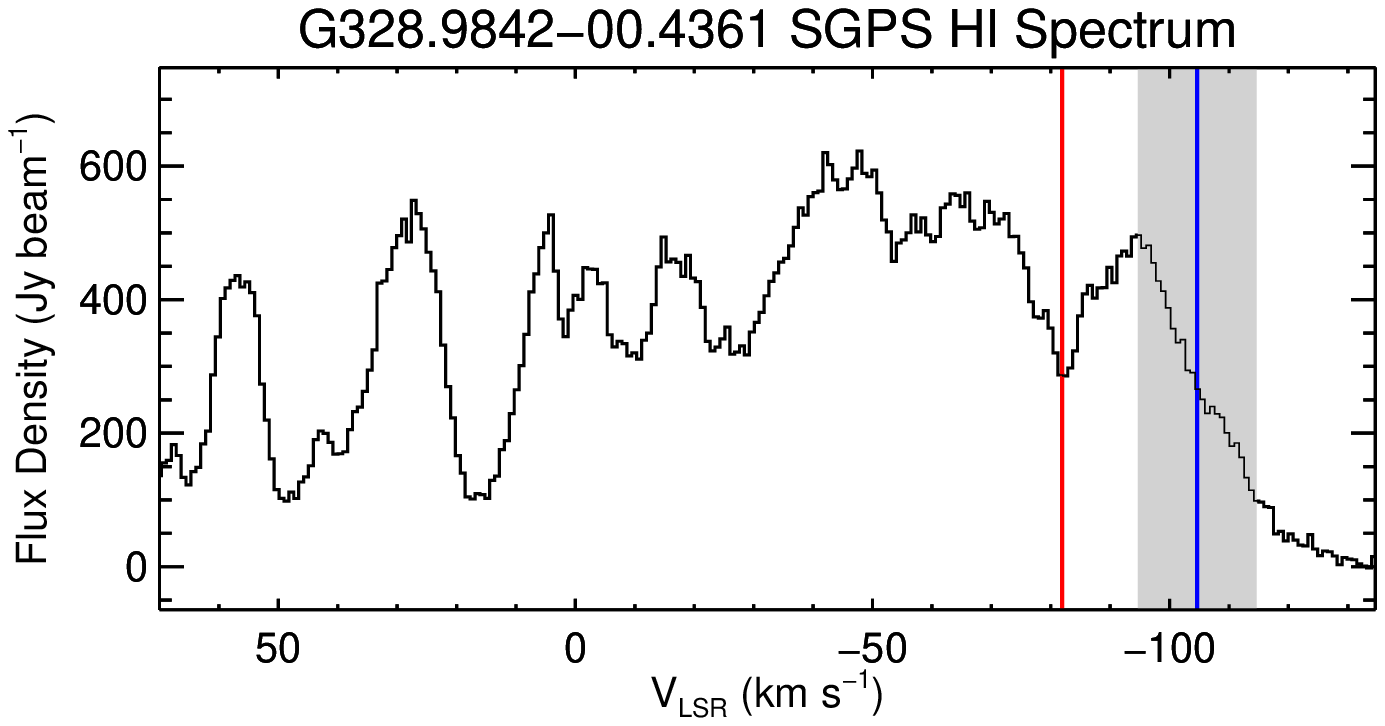}
\includegraphics[width=0.49\textwidth, trim= 0 0 0 0]{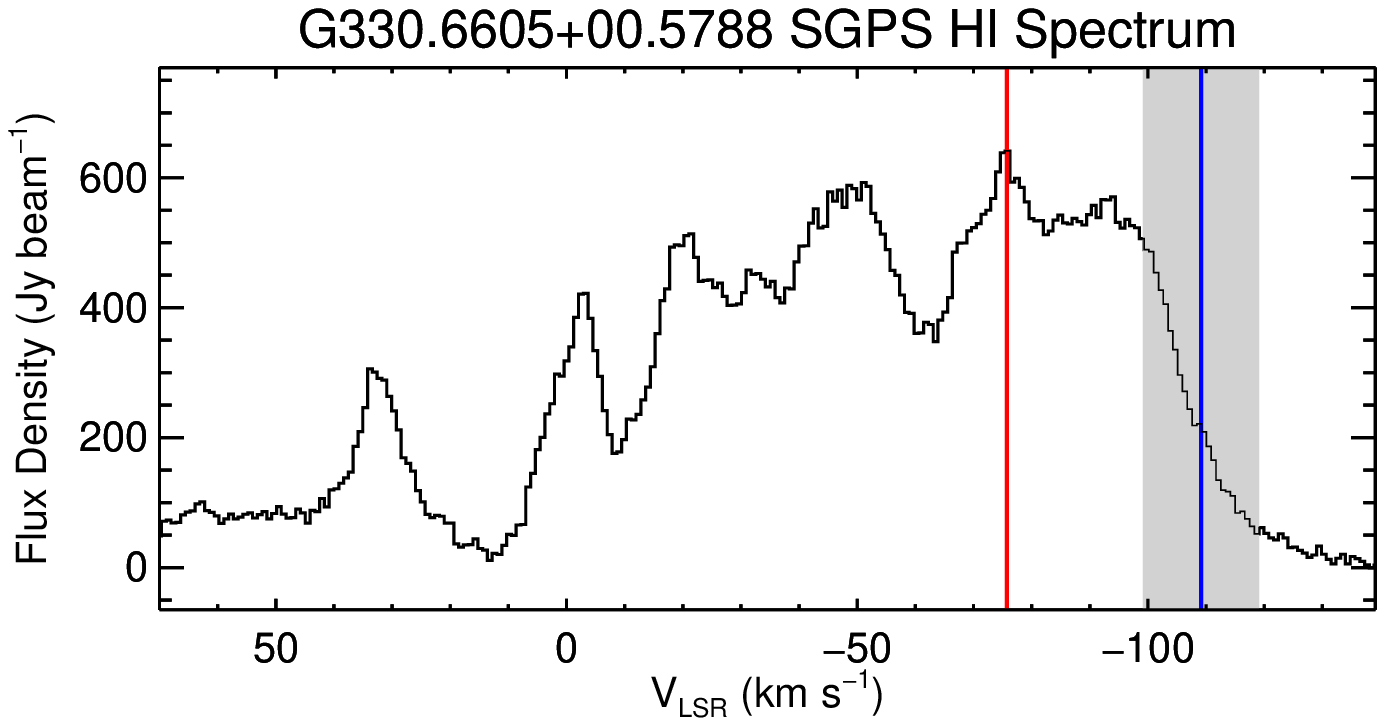}

\includegraphics[width=0.49\textwidth, trim= 0 0 0 0]{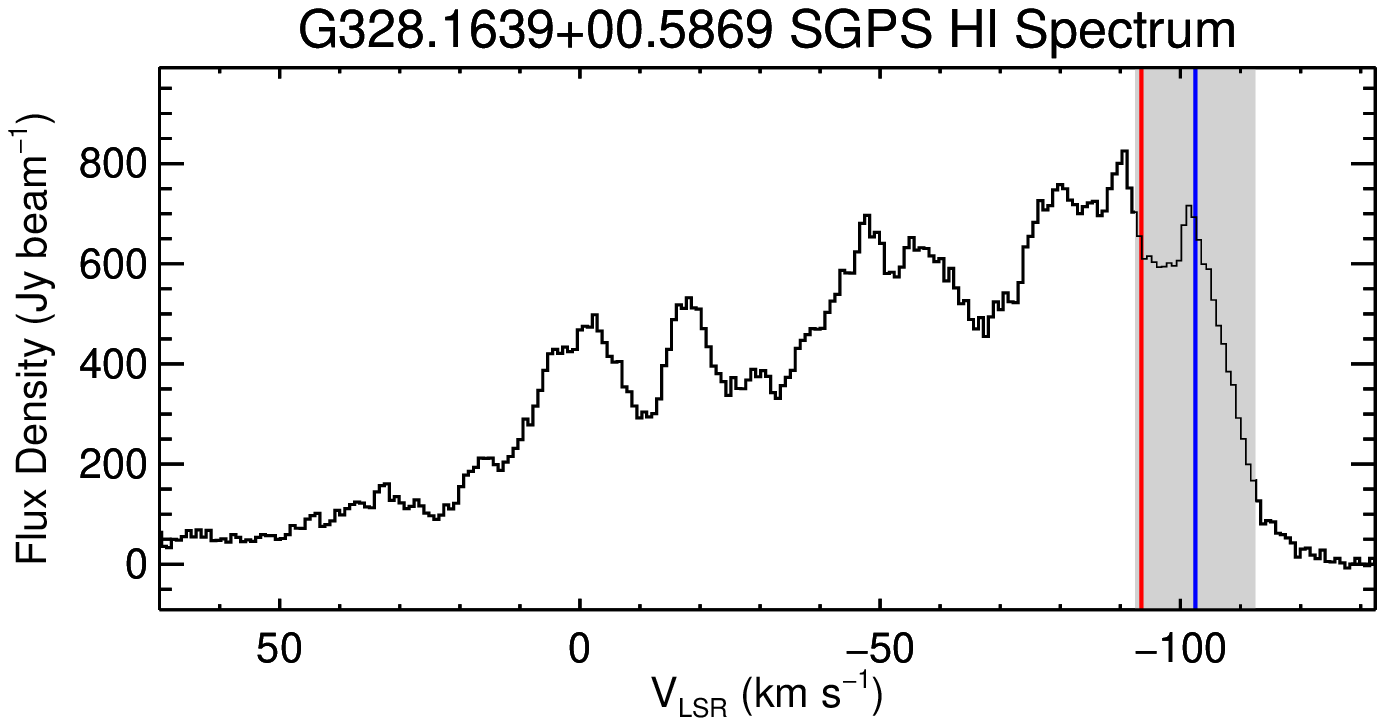}

\caption{\label{fig:HISA_examples} Examples of SGPS
  \hi\ spectra towards the embedded RMS sources located within the Solar circle for which a reliable distance was not available in the literature. The source
  velocity ($v_{\rm{s}}$) and the velocity of the tangent point ($v_{\rm{t}}$) are shown by the red and blue vertical lines, respectively. The grey vertical band covers the velocity region 10\,\kms\ either side of the tangent velocity and is provided an easy way to identify sources placed at the tangent position. In the upper, middle and lower panel we present and example of a source located at the near, far and tangent positions.} 

\end{center}
\end{figure}

Distances are critical to estimating source luminosities, which allow us to discriminate between genuinely massive stars and nearby lower-mass stars and to determine their Galactic distribution. Deriving distances to our sample of MYSOs and \hii\ regions has been an ongoing process, with results having previously been presented for incomplete samples located primarily in the first and fourth quadrants (\citealt{urquhart2011a} and \citealt{urquhart2012}, respectively). 

Maser parallax derived distances are the most reliable methods to determine distances and these are becoming available for an increasing number of star-forming regions (i.e., \citealt{reid2009}). At present, these tend to be localised to the relatively nearby parts of the first and second quadrants of the Galaxy, and only provide distances to a small fraction of our sample (no more than a few per\,cent). Spectrophotometric measurements of \hii-region-exciting stars or associated clusters are the next most reliable method, however, this can produce incorrect distances if the star is incorrectly typed (see \citealt{clark2009} for an example). We have conducted a comprehensive review of the literature and have adopted both parallax and spectrophotometric distances where available, and will continue to update source distances as and when new measurements become available. 

For the remaining sources we have estimated kinematic distances using the Galactic rotation model derived by \citet[][circular rotation speed $\theta_0= 254$\,\kms\ and distance to Galactic centre $R_0 = 8.4$\,kpc]{reid2009} and the radial velocities discussed in the previous section. Kinematic distances have associated uncertainties of order $\pm$1\,kpc (allowing for $\pm$7\,\kms\ error in the velocity due to streaming motions). We also note that kinematic distances have been found to deviate significantly from parallax and spectroscopically derived distances (e.g., \citealt{moises2011}). Although these distance anomalies can dramatically affect the derived properties of individual sources it will not impact of the results drawn from the whole sample.  Of greater concern is that within the Solar circle there are two solutions for each radial velocity (\textit{kinematic distance ambiguity}, or KDA). These distances are equally-spaced on either side of the tangent position, and are generally referred to as the \textit{near} and \textit{far} distances. Since the majority of our sources are located within the Solar circle, this distance ambiguity affects $\sim$80\,per\,cent of our sample. 

\setlength{\tabcolsep}{6pt}

\begin{table*}

\begin{center}
\caption{Results of the \hi\ self-absorption analysis.}
\label{tbl:hisa_results}
\begin{minipage}{\linewidth}
\begin{tabular}{l.....c...}
\hline
\hline

\multirow{2}{24mm}{MSX Name}	&	\multicolumn{1}{c}{$\ell$}  		&\multicolumn{1}{c}{$b$}  	& \multicolumn{1}{c}{\vlsr}	& \multicolumn{1}{c}{Near}& \multicolumn{1}{c}{Far} &	\multirow{2}{6mm}{KDS$^a$}	&  \multicolumn{1}{c}{Distance} & \multicolumn{1}{c}{$R_{\rm{GC}}$}&\multicolumn{1}{c}{$z$} \\
&	\multicolumn{1}{c}{(\degr)} 	& \multicolumn{1}{c}{(\degr)} 	&\multicolumn{1}{c}{(\kms)} & \multicolumn{1}{c}{(kpc)} & \multicolumn{1}{c}{(kpc)} 		& \multicolumn{1}{c}{} &\multicolumn{1}{c}{(kpc) } 	&\multicolumn{1}{c}{(kpc)} 	& \multicolumn{1}{c}{(pc)}   \\  

\hline
G327.9205+00.0921	&	327.9210	&	0.0921	&	-49.5	&	3.2	&	11.1	&	F	&	11.1	&	5.9	&	17.8	\\
G328.5487+00.2717	&	328.5490	&	0.2717	&	-60.5	&	3.7	&	10.6	&	N	&	3.7	&	5.6	&	17.6	\\
G327.8097$-$00.6339	&	327.8100	&	-0.6339	&	-46.7	&	3.0	&	11.2	&	N	&	3.0	&	6.1	&	-33.6	\\
G328.2275$-$00.2714	&	328.2280	&	-0.2714	&	-99.5	&	5.8	&	8.5	&	F	&	7.1	&	4.4	&	-33.8	\\
G328.9480+00.5709	&	328.9480	&	0.5709	&	-93.9	&	5.3	&	9.0	&	F	&	7.2	&	4.3	&	71.7	\\
G328.9580+00.5671	&	328.9580	&	0.5671	&	-93.5	&	5.3	&	9.1	&	F	&	7.2	&	4.3	&	71.2	\\
G329.2713+00.1147	&	329.2710	&	0.1147	&	-76.9	&	4.5	&	9.9	&	N	&	4.5	&	5.1	&	9.0	\\
G329.3371+00.1469	&	329.3370	&	0.1469	&	-107.1	&	6.1	&	8.3	&	F	&	7.2	&	4.3	&	18.5	\\
G329.4720+00.2143	&	329.4720	&	0.2143	&	-101.5	&	5.7	&	8.7	&	F	&	7.2	&	4.3	&	27.1	\\
G329.4579+00.1724	&	329.4580	&	0.1724	&	-103.2	&	5.8	&	8.6	&	F	&	7.2	&	4.3	&	21.8	\\
G328.9842$-$00.4361	&	328.9840	&	-0.4361	&	-81.9	&	4.7	&	9.7	&	N	&	4.7	&	5.0	&	-36.1	\\
G329.6098+00.1139	&	329.6100	&	0.1139	&	-63.9	&	3.9	&	10.6	&	N?	&	3.9	&	5.4	&	7.7	\\
G329.4211$-$00.1631	&	329.4210	&	-0.1631	&	-76.8	&	4.5	&	10.0	&	N	&	4.5	&	5.1	&	-12.8	\\
G329.8145+00.1411	&	329.8150	&	0.1411	&	-85.0	&	4.9	&	9.7	&	N	&	4.9	&	4.8	&	12.0	\\
G329.3402$-$00.6436	&	329.3400	&	-0.6435	&	-74.2	&	4.4	&	10.1	&	F	&	10.1	&	5.2	&	-113.2	\\
G330.6605+00.5788	&	330.6600	&	0.5788	&	-75.7	&	4.4	&	10.2	&	F	&	10.2	&	5.0	&	103.2	\\
G330.2923+00.0010	&	330.2920	&	0.0010	&	-64.3	&	3.9	&	10.7	&	N?	&	3.9	&	5.4	&	0.1	\\
G331.0890+00.0163	&	331.0890	&	0.0163	&	-95.6	&	5.3	&	9.4	&	N	&	5.3	&	4.5	&	1.5	\\
G331.0931$-$00.1303	&	331.0930	&	-0.1303	&	-64.0	&	3.9	&	10.8	&	F?	&	10.8	&	5.3	&	-24.6	\\
G330.9288$-$00.4070	&	330.9290	&	-0.4070	&	-41.6	&	2.8	&	11.9	&	F	&	11.9	&	6.1	&	-84.2	\\
\hline
\end{tabular}

\footnotetext[1]{Kinematic Distance Solution.  `N' indicates that the source was determined to be at the \textit{near} location, while an `F' reflects the \textit{far} position. Where a distance solution is considered less reliable we append '?' to the  given allocation.}

\end{minipage}
\end{center}
Notes: A small portion of the data is provided here, the full table is only  available in electronic form at the CDS via anonymous ftp to cdsarc.u-strasbg.fr (130.79.125.5) or via http://cdsweb.u-strasbg.fr/cgi-bin/qcat?J/MNRAS/.

\end{table*}
\setlength{\tabcolsep}{6pt}

We  reduce the number of ambiguities that need to be resolved by applying two initial cuts. First, we place sources  with velocities within 10\,\kms\ of the tangent velocity at the tangent distance, since the error in the distance is comparable to the difference between the near/far distance and the tangent distance. Second, we place sources at the near distance if a far allocation would result in an unrealistically large displacement from the Galactic mid-plane for a star formation region (i.e., $z$ $>$ 120\,pc; 4 $\times$ scaleheight for O-type stars; \citealt{reed2000}). These two steps reduce the number of KDAs that need to be resolved by {\color{black}$\sim$100}; however, the vast majority still need to be addressed.

There are a number of ways to resolve these kinematic distance ambiguities: \hi\ absorption against a continuum \citep[e.g.,][]{fish2003, kolpak2003,anderson2009a,roman2009,urquhart2012}, \hi\ self-absorption \citep[e.g.,][]{jackson2003,roman2009,green2011b}, and matching sources with infrared dark clouds \citep[e.g.,][]{dunham2011b, ellsworth2013}. These techniques are well documented in the literature and so will not be described here in detail.  Many of these studies 
have sources in common with our sample, and we have compared the positions and velocities (i.e., $\ell bv$ parameter space) of these and adopted the given distance solution where a match is found. In cases where the distance solutions differ we have made an independent assessment of the available data to determine the most likely distance and have favoured the allocation based on the most reliable data (i.e., the highest resolution and/or data with the highest signal to noise ratio). In cases where there is no clear distinction we do not resolve the ambiguity (but may do so in the future as more data become available). 

Since all of the bright \hii\ regions have been included in previous studies using the \hi\ absorption against a continuum method, the $\sim$200 sources for which the ambiguity has not been resolved are either radio-quiet or the radio emission is too weak for absorption to be detected in the \hi\ surveys available. We have therefore extracted \hi\ spectra from the Southern Galactic Plane Survey (SGPS; \citealt{mcclure2005}) and the VLA Galactic Plane Survey (VGPS; \citealt{stil2006}) for all these remaining objects. These have been inspected for absorption coincident with the source velocity, indicating the near distance, otherwise the far distance is considered more likely. Additionally, we inspected the GLIMPSE and MIPSGAL images to correlate sources located at the near distance with IRDCs in an effort to confirm the \hi\ results. If a correlation was found the near distance was assigned, if no correlation was found no distance was assigned. Fig.\,\ref{fig:HISA_examples} shows some examples of these distance allocations and in Table\,\ref{tbl:hisa_results} we present the distance solutions for all sources examined. 

Using a combination of parallax and spectroscopic distances found in the literature, and KDA solutions drawn both from the literature and derived from our own analysis, we are able to assign distances to $\sim${\color{black}1650} of the embedded RMS sources. We have been unable to assign a distance to approximately 100 sources, however, this corresponds to $\sim$7\,per\,cent of the embedded RMS population.

\subsection{Bolometric luminosities and survey completeness}

\begin{figure}
\begin{center}

\includegraphics[width=0.45\textwidth, trim= 0 0 0 0]{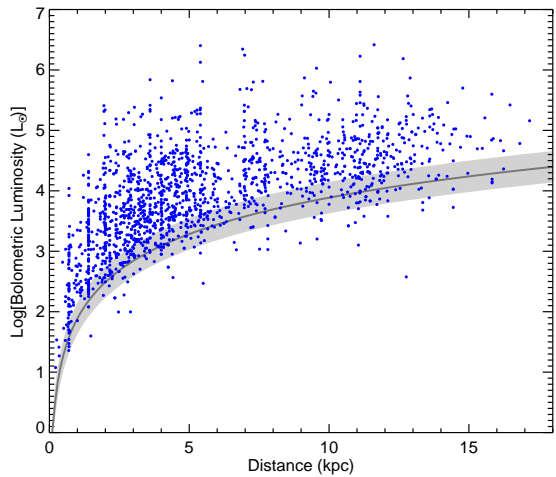}

\caption{\label{fig:rms_luminosity} The luminosity distribution as a function of heliocentric distance. The dark line and light grey shaded region indicates the limiting sensitivity of the MSX 21\,\mum\ band and its associated uncertainty. Where multiple sources have been identified within the MSX beam, the luminosity has been apportioned, resulting in some sources being located below the sensitivity limit.}

\end{center}
\end{figure}

Bolometric luminosities have been estimated for the majority of sources using the distances discussed in the previous section and the integrated fluxes determined from model fits to the spectral energy distributions (SEDs) presented by \citet{mottram2011b}. For very bright sources and sources located in complicated regions it is not always possible to obtain a sufficient number of photometric points to adequately constrain the SED; this is primarily due to saturation of the mid- and far-infrared images and difficulties subtracting the diffuse background emission (\citealt{mottram2010}). In these cases the luminosities have been estimated by simply scaling the MSX 21\,$\umu$m flux (see \citealt{mottram2011a} for details). 

The RMS luminosities have been previously reported by \citet{mottram2011b}, who used a less complete RMS subset to investigate the luminosity functions of MYSO and \hii\ regions and determine statistical lifetimes of these stages. Rather than reproducing this work, we aim simply to use the bolometric luminosities to estimate the survey's completeness threshold for the full sample of embedded RMS sources and to investigate the distribution of massive stars across the Galaxy.\footnote{The luminosities presented by \citet{mottram2011b} were estimated using kinematic distances derived from the \citet{brand1993} rotation curve and so have been rescales to the kinematic distances derived from the \citet{reid2009} rotation model.}

In  Fig.\,\ref{fig:rms_luminosity} we show the distribution of YSO and \hii\ region bolometric luminosities as a function of heliocentric distance. This plot suggests that the sample is complete to young embedded sources with bolometric luminosities over $\sim$$2\times10^4$\,\lsun\ to a distance of $\sim$18\,kpc.   While the bolometric luminosities for individual sources can be found in \citet{lumsden2013}, we present the total bolometric luminosities of star forming complexes in Table\,\ref{tbl:complex_parameters}. These have been estimated by integrating the bolometric luminosities of the embedded massive stellar population associated with each complex.

\begin{figure*}
\begin{center}
\includegraphics[width=.9\textwidth, trim= 50 0 50 0]{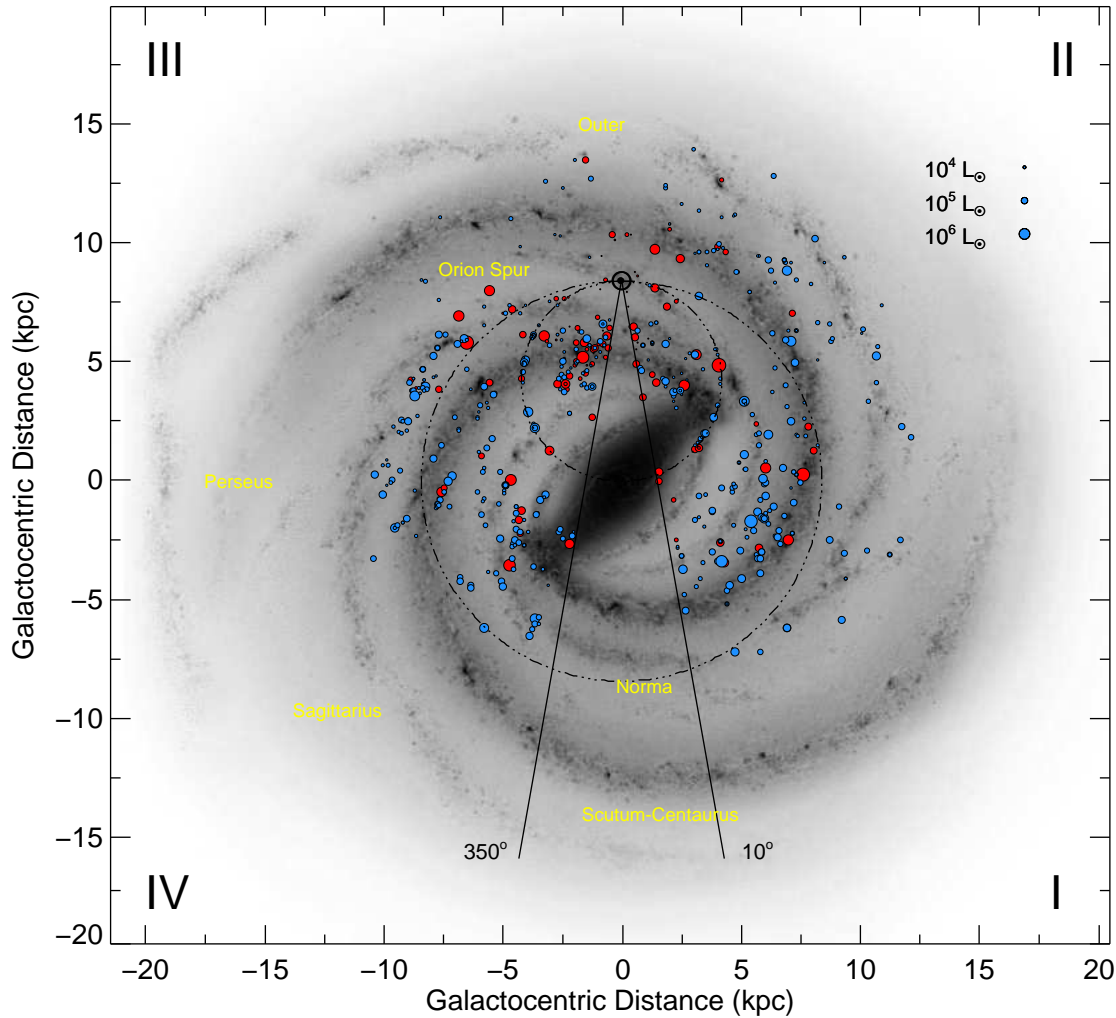}

\caption{\label{fig:galactic_mass_radius_distribution}Galactic distribution of all MYSOs and \hii\ regions with bolometric luminosities greater than $10^4$\,\lsun. We show the kinematic positions of the complexes and individual sources as red and blue circles, respectively.  The sizes of the markers give an indication of their luminosity, as depicted in the upper right corner. The background image is a sketch of the Galaxy produced by Robert Hurt of the Spitzer Science Center in consultation with Robert Benjamin. The position of the Sun is shown by the small circle above the Galactic centre. The two solid black lines enclose the Galactic centre region that was excluded from the RMS surveys due to problems with source confusion and distance determination. The smaller of the two black dot-dashed circles represent the locus of tangent points, while the larger circle shows the radius of the Solar circle.} 

\end{center}
\end{figure*} 

\section{Galactic Structure}
\label{sect:gal_structure}

With the distances on hand we are in a position to examine the 3-dimensional distribution of this sample of young massive stars. In Fig.\,\ref{fig:galactic_mass_radius_distribution} we present a top-down view of the Milky Way showing the distribution of MYSOs and \hii\ regions with respect to the known large-scale structures of the Galaxy. The background image shown in this figure is an artist's conceptual image of what the Galaxy might look like if viewed from above, and incorporates all that is currently known of the structure of our Galaxy, including the 3.1-3.5\,kpc Galactic bar at an angle of 20\degr\ with respect to the Galactic centre-Sun axis (\citealt{binney1991, blitz1991, dwek1995}), a second non-axisymmetric structure referred to as the ``Long bar'' (\citealt{hammersley2000}) with a Galactic radius of 4.4$\pm$0.5\,kpc at an angle of 44$\pm$10\degr\ (\citealt{benjamin2008}), the Near and Far 3-kpc arms, and the four principle arms (i.e., Norma, Sagittarius, Perseus and Scutum-Centaurus arms). The location of the spiral arms is based on the \citet{georgelin1976} model but has been modified to take account of recent maser parallax distances and updated directions for the spiral arm tangents (\citealt{dame2001}).  The RMS survey excluded the innermost 20\degr\ of the Galactic longitude (i.e., 350\degr $< \ell <$ 10\degr): we are therefore not sensitive to any star formation taking place within $\sim$3\,kpc of the Galactic centre and any features of Galactic structure within this radius.

The distribution of the young embedded population with respect to the expected positions of the spiral arms given from the models suggests that they are correlated, but quantifying this correlation and estimating its significance is non-trivial. The correlation between the spiral arms and the RMS complexes appears slightly stronger than for the isolated sources in that there are far fewer found to be located in the inter-arm regions (cf \citealt{stark2006}). The reason for this is that the more reliable maser parallax and spectrophotometric distances are available for many of the complexes, and for the others, the higher number of associated RMS sources is likely to provide a better estimates of their systemic radial velocities and thus kinematic distances (cf \citealt{russeil2003}). Individual sources are by comparison more poorly constrained. The source density of the inner parts of the spiral arms ($R_{\rm{GC}} < 10$\,kpc) appears to be roughly uniform, which suggests a similar level of star formation is taking place within them. This is in contrast with the result of near- and mid-infrared source counts presented by \citet{benjamin2008}. These authors found enhancements of the source counts towards the Scutum and Centaurus tangents, but not towards the Norma or Sagittarius tangents. This led them to speculate that the Galaxy has two principal arms (i.e., Perseus and Scutum-Centaurus arms) with the Norma or Sagittarius perhaps being optically visible arms that are not associated with any enhancement in the old stellar disk. The emphasised Perseus and Scutum-Centarurus arms portrayed in the background image used in Fig.\,\ref{fig:galactic_mass_radius_distribution} reflect this \citep{benjamin2008,churchwell2009}.  

\begin{figure}
\begin{center}

\includegraphics[width=0.45\textwidth, trim= 0 0 0 0]{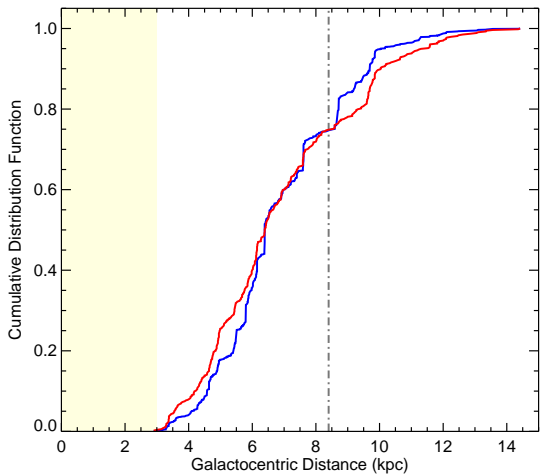}

\caption{\label{fig:rgc_cdf_diagram} Cumulative distribution function of the RMS source counts (red) and integrated bolometric luminosities (blue) of all MYSOs and \hii\ regions above $2\times10^4$\lsun\ as a function of Galactocentric radius. The shaded region shows the Galactocentric distance region excluded from the RMS survey, while the vertical dashed line indicates the distance of the Solar circle. } 

\end{center}
\end{figure}

Comparison of the number of northern and southern RMS sources above the completeness limit shows a similar number of sources (i.e., $47\pm4.6$\,per\,cent and $53\pm4.0$\,per\,cent for the northern and southern Galactic plane, respectively). We also find similar proportions of the northern and southern samples inside and outside the Solar circle ($\sim$75\,per\,cent and $\sim$25\,per\,cent, respectively). The proportional distribution of the RMS sources between the northern and southern Galactic plane and inside and outside the Solar circle are similar to that reported by \citet{bronfman2000} from a study of $\sim$750 $IRAS$ selected candidate \uchii\ regions (67\,per\,cent and 33\,per\,cent inside and outside the Solar circle). If we look at the integrated bolometric luminosities of the sources inside and outside the Solar circle we also find the same fractional distribution as the source counts; this is nicely illustrated in the cumulative distribution function of both of these parameters in Fig.\,\ref{fig:rgc_cdf_diagram}. The fraction of the total bolometric luminosity of sources within the Solar circle is slightly lower than the value of 81\,per\,cent reported by \citet{bronfman2000}, however, it is still consistent within the errors.

The similarity of the distributions of the total luminosity and source counts both inside and outside of the Solar circle would suggest that the mean source luminosity is also likely to be broadly similar.  Furthermore, given that the proportion of molecular gas inside and outside the Solar circle is very similar to the luminosity and source counts it follows that the average star formation efficiencies are also comparable. This is consistent with the results of \citet{snell2002} that the star formation efficiency for the population of molecular clouds in the outer Galaxy (as estimated from the clouds' $L_{\mathrm{FIR}}/M$ ratio) is similar to that found for the inner Galaxy population. The clouds in the outer Galaxy are as active in forming massive stars, despite the fact that the clouds are typically 1-2 orders of magnitude less massive than those in the inner Galaxy.  The difference in star formation rates between inner and outer regions, then, is principally a function of the molecular cloud formation rate, not of star formation $within$ the molecular clouds.  Several factors are thought to contribute to the comparative lack of molecular clouds, including lower metallicity, gas density, and ambient pressure \citep{leroy2008,snell2002,schruba2011}.

\subsection{Surface density of massive star formation} 

\begin{figure*}
\begin{center}

\includegraphics[width=0.45\textwidth, trim= 0 0 0 0]{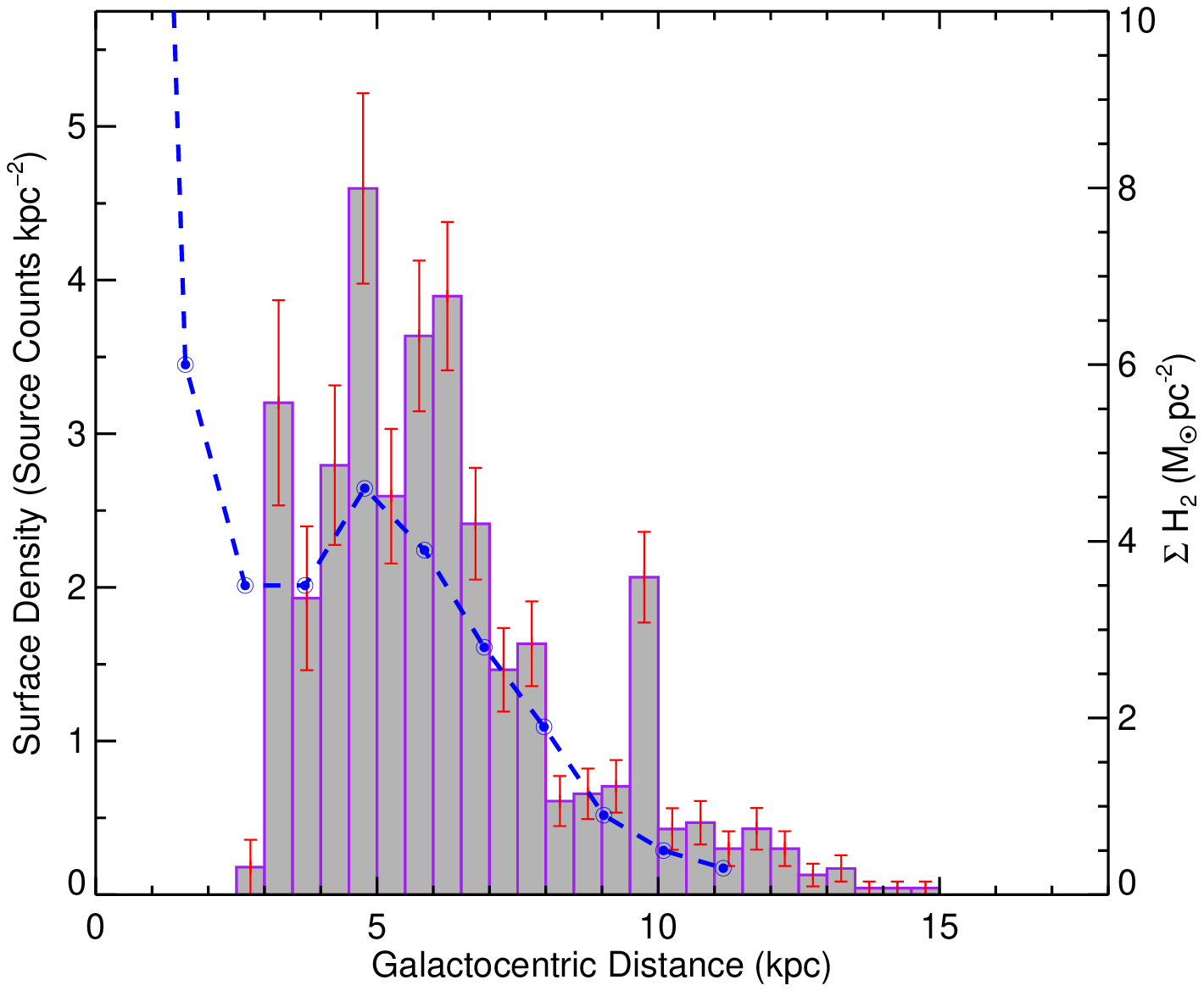}
\includegraphics[width=0.45\textwidth, trim= 0 0 0 0]{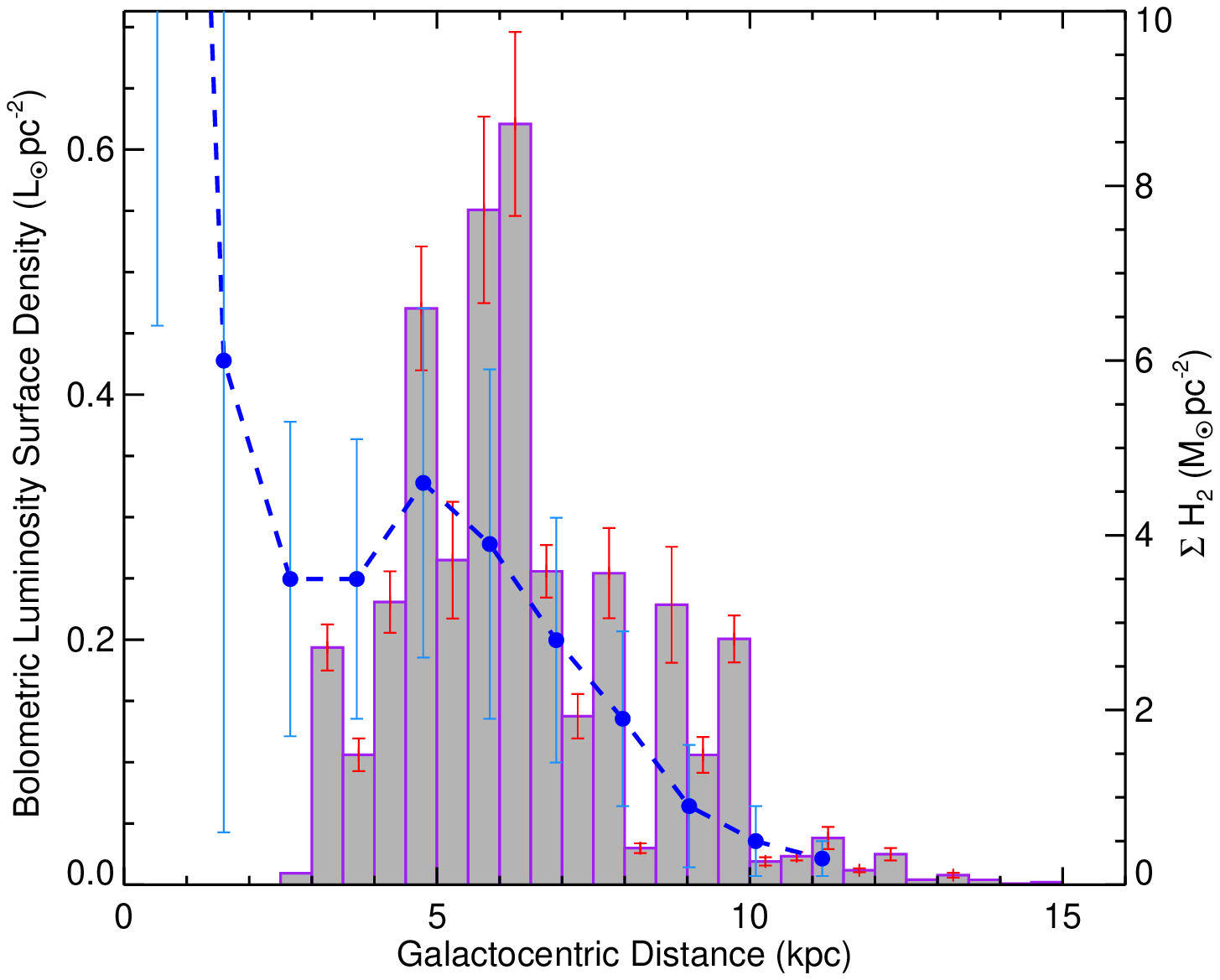}

\includegraphics[width=0.45\textwidth, trim= 0 0 0 0]{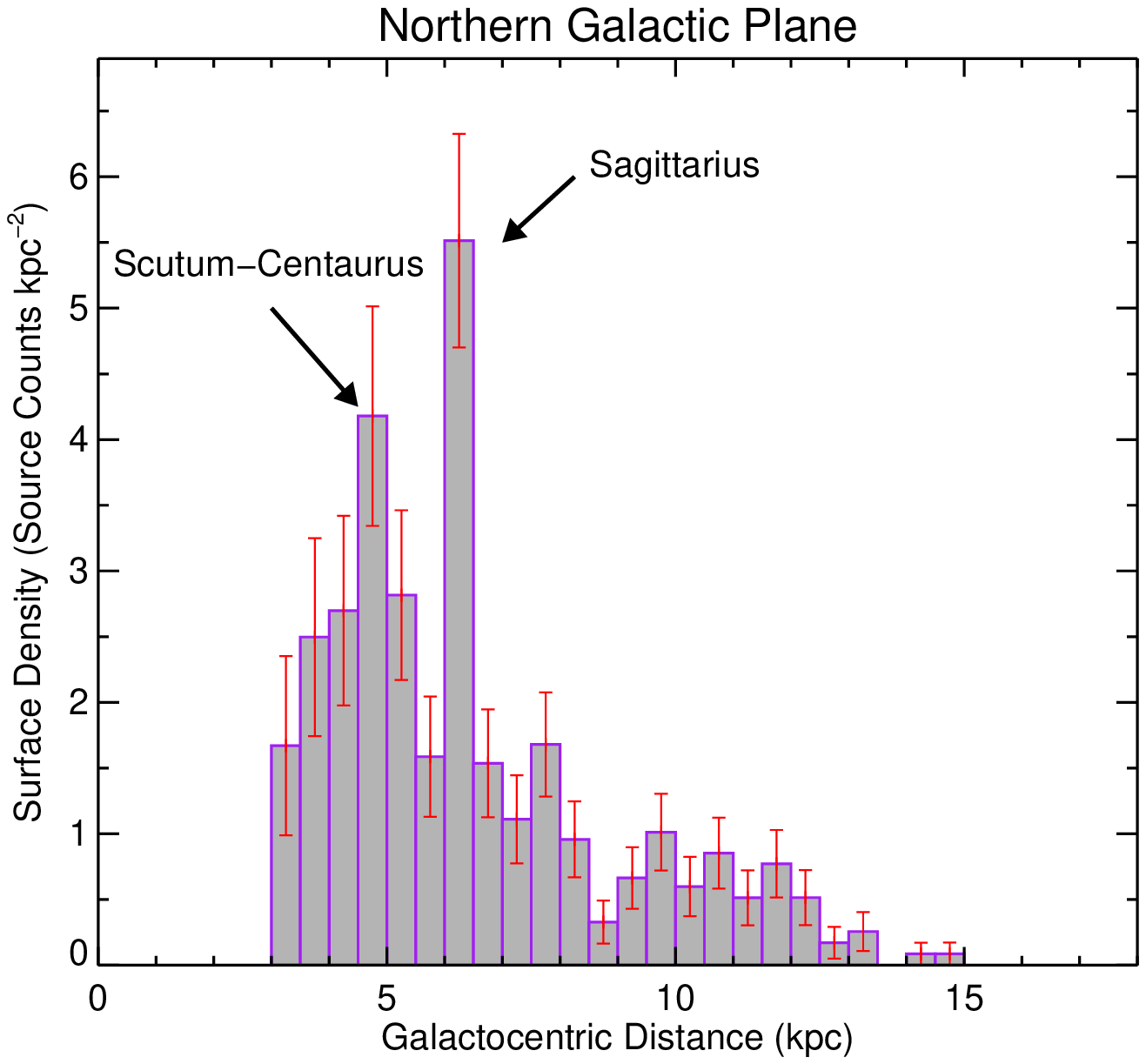}
\includegraphics[width=0.45\textwidth, trim= 0 0 0 0]{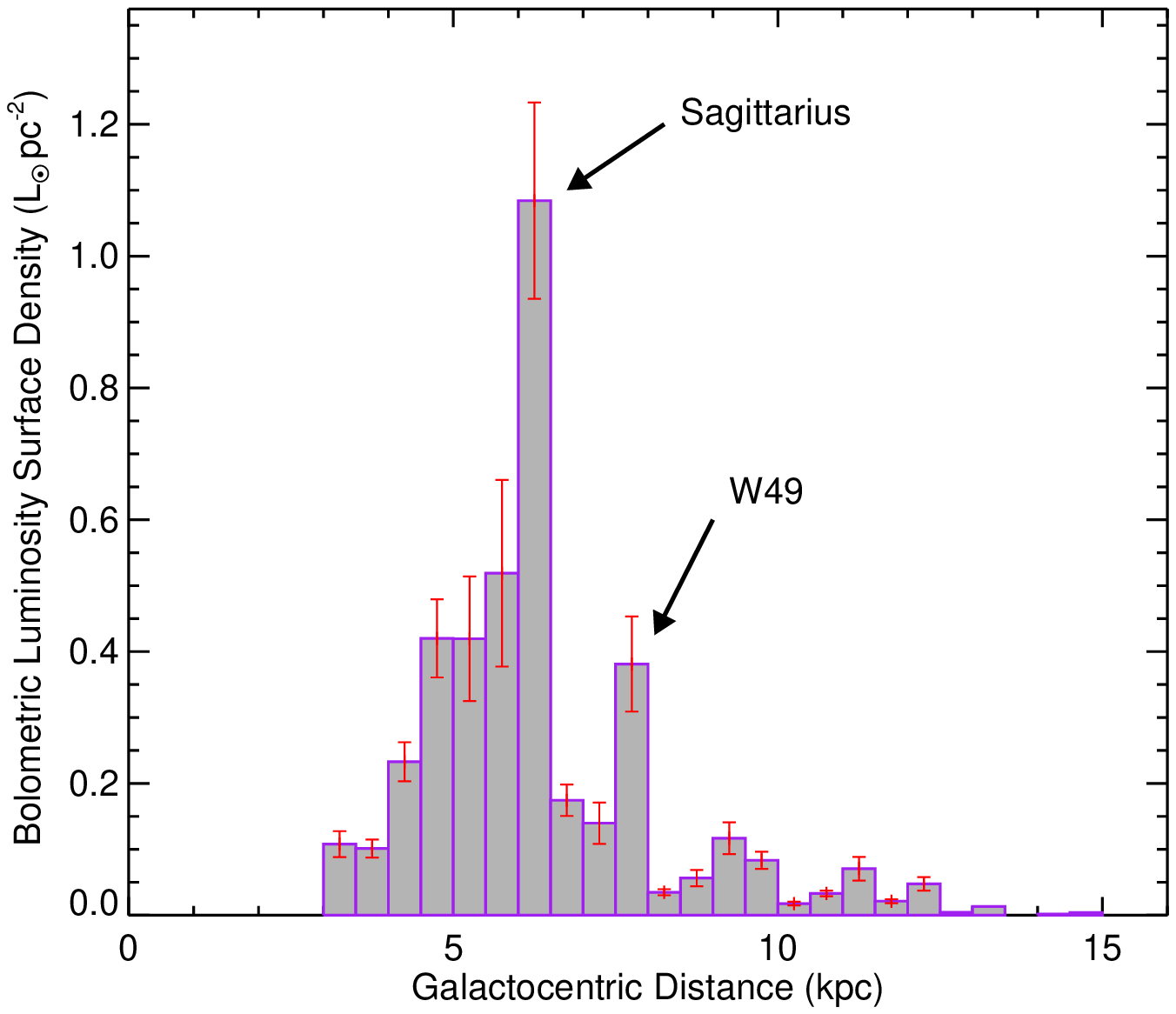}

\includegraphics[width=0.45\textwidth, trim= 0 0 0 0]{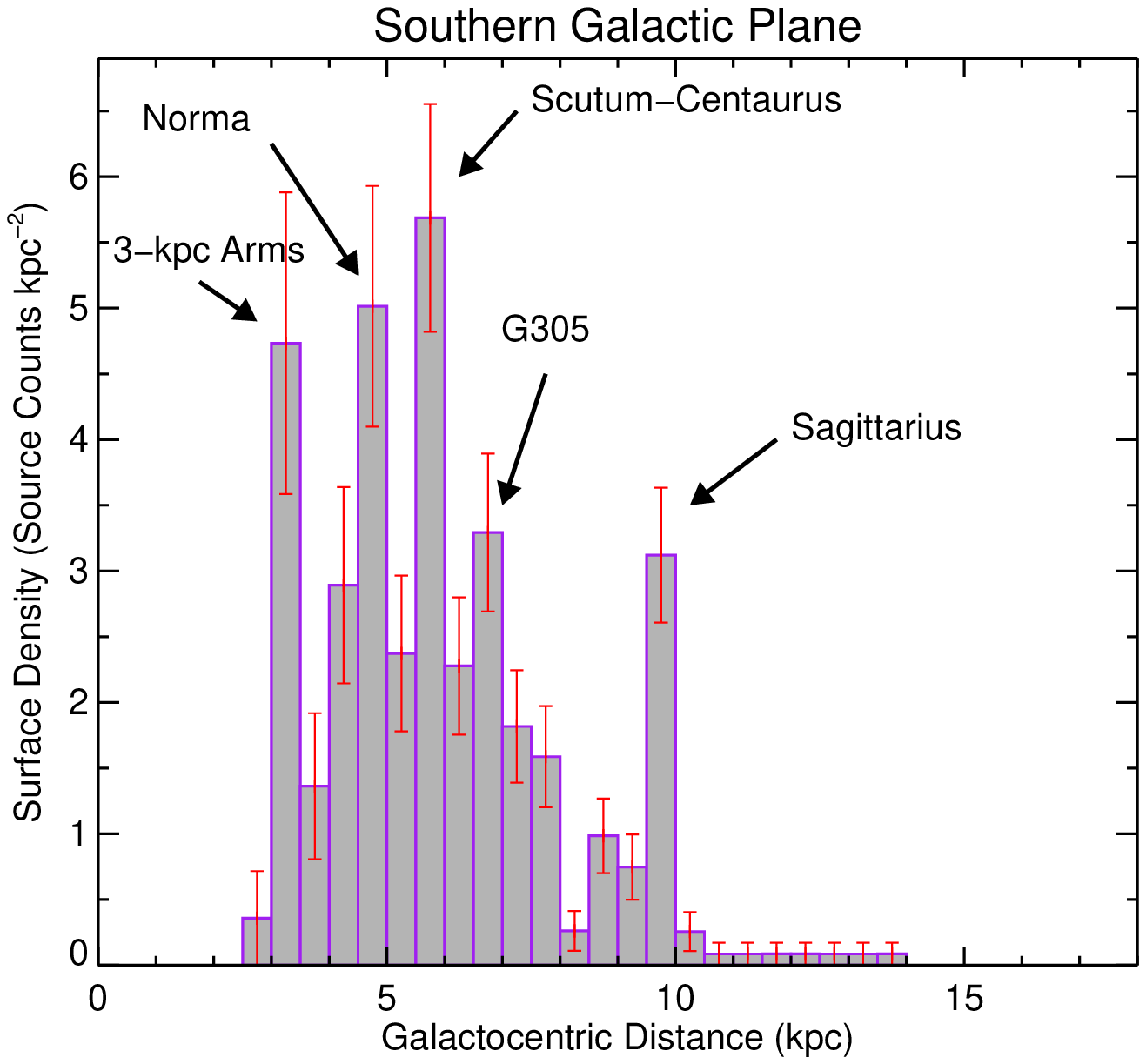}
\includegraphics[width=0.45\textwidth, trim= 0 0 0 0]{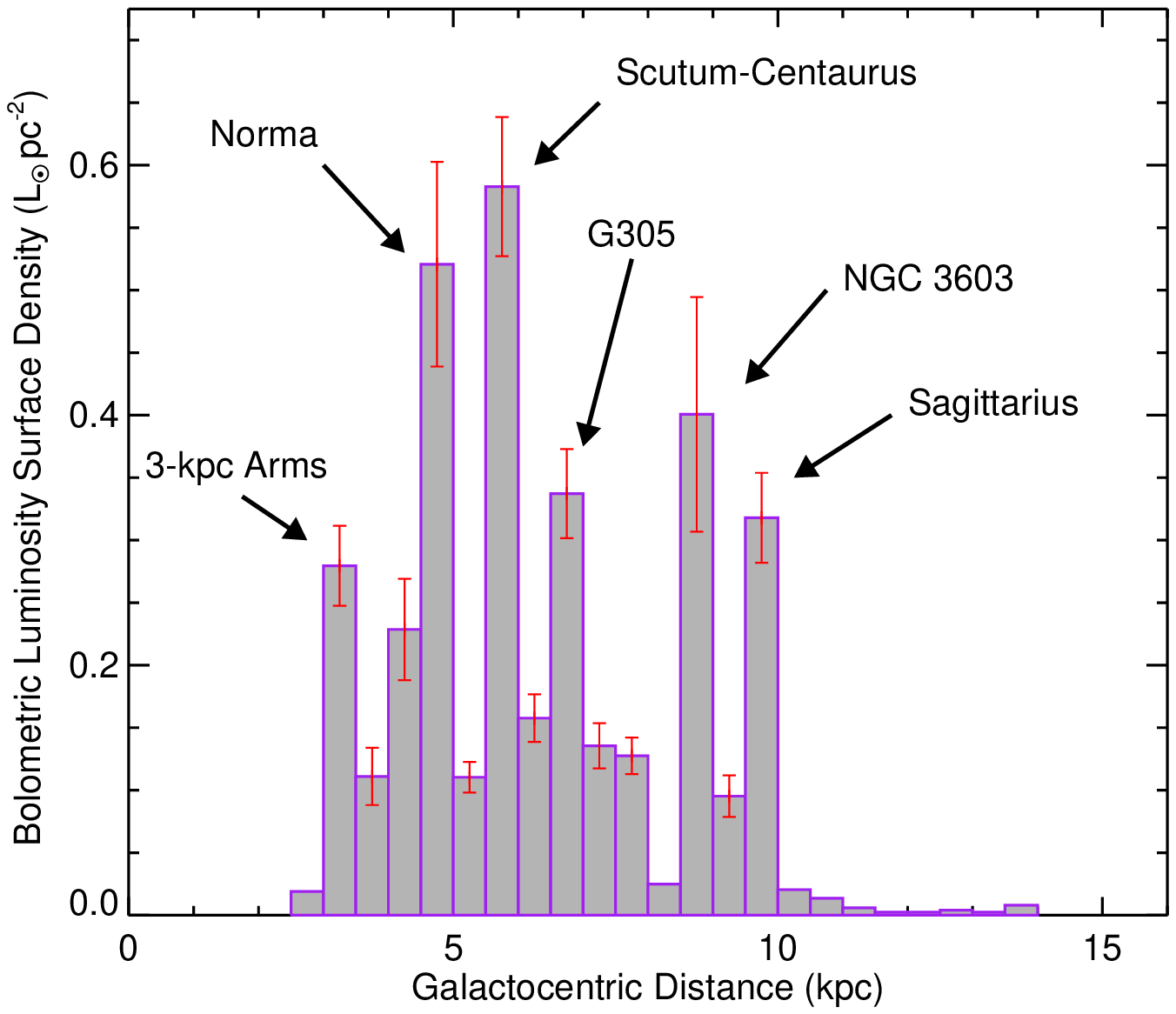}

\caption{\label{fig:rms_rgc_distribution}  Source  and bolometric luminosity surface density distributions of the population of RMS sources above the sample completeness limit ($2\times 10^4$\,\lsun).  The upper panel shows the distribution of the whole sample, while the middle and lower panels show the distribution of the northern and southern samples, respectively. The bin size is 0.5\,kpc and the uncertainties in the left panels are derived from \poi\ statistics, while the those in the right panels are estimated assuming a 40\,per\,cent error in the derived source luminosities. The blue circles and connecting dashed blue line indicates the H$_2$ gas surface density taken from \citet[][see righthand $y$-axis in upper panel for values]{nakanishi2006}. The light blue vertical lines indicate the errors associated with the H$_2$ gas surface density points.} 

\end{center}
\end{figure*}

In the upper panels of Fig.\,\ref{fig:rms_rgc_distribution} we present the massive star formation and bolometric luminosity surface density distributions as a function of Galactic radius. These plots include only sources with luminosities above the completeness threshold. However, sources above this limit dominate the total luminosity, and inclusion of the lower luminosity sources has very little impact on the measured values or the overall structure of the distribution. This has been produced in the standard way by dividing the total number of sources in each annulus by its area.  We have also applied a heliocentric radius limit of 17\,kpc to avoid including areas that are effectively outside the regions the RMS survey is sensitive to and the area of the wedge towards the inner Galaxy excluded from the RMS survey (i.e., $350\degr < \ell < 10\degr$). The overall distributions are both highly structured showing numerous peaks, many at similar radii. In the middle and lower panels of this figure we separate these distributions into the northern and southern populations, respectively, and identify the spiral arms that are coincident with the observed peaks in the first and fourth quadrants.

\subsubsection{Source counts}

Peaks are seen in the RMS source distribution for the whole sample at $\sim$3, 5, 6 and 10\,kpc. The 5 and 6\,kpc peaks are coincident with the segments of the  \Scu\ and \Sag\ arms located in the northern Galactic plane, but no peaks are found towards the  \Per\ and \Nor\ arms located in this part of the Galaxy. Both of these arms, however, extend into the outer Galaxy and are spread over a larger range of Galactic radii than the other arms (6-10\,kpc for and 8-13\,kpc for the \Per\ and \Nor\ arms, respectively), thus smearing their Galactocentric distribution. All four distinct peaks seen in the upper left panel of Fig.\,\ref{fig:rms_rgc_distribution} (i.e., $\sim$3, 5, 6 and 10\,kpc) are seen in the southern Galactic plane distribution; these can be attributed to the  Near and Far 3-kpc arms, the \Nor, \Scu\ and  \Sag\ arms, respectively.

All of the peaks located within the Solar circle have been previously reported for RMS subsamples located in the first and fourth quadrant (i.e., \citealt{urquhart2011a} and  \citealt{urquhart2012}, respectively). The overall distribution, as well as that of the northern and southern Galactic plane samples, is also very similar to those presented by  \citet{bronfman2000}. Similar features have also been reported in the Galactocentric distribution of methanol masers \citep{green2011b}, \hii\ regions \citep{anderson2009a,paladini2004}, thermal dust emission \citep{dunham2011b}  and molecular gas \citep{rathborne2009}. The correlation of these observed peaks in the Galactocentric distribution, and between the radial velocities and Galactic longitude of the RMS sources and the spiral arms models (i.e., Fig.\,\ref{fig:lv_distribution}) is consistent with four-armed models of the Galaxy (e.g., \citealt{georgelin1976, nakanishi2006}). 

\begin{figure*}
\begin{center}

\includegraphics[width=0.99\textwidth, trim= 0 0 0 0]{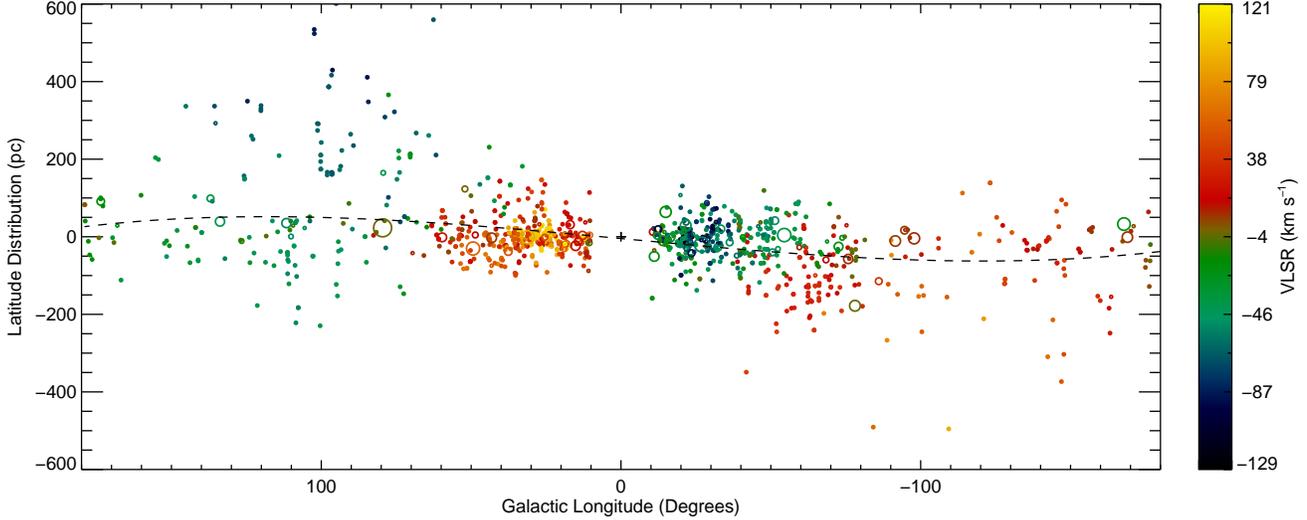}

\caption{\label{fig:l_z_distribution} Distance from the Galactic mid-plane (\zo) for the embedded RMS source population is shown as a function of Galactic longitude. The complexes are shown as open circles (larger symbol sizes indicate a greater number of members), while individual sources are shown as filled circles. The symbol colours give the source or systematic velocity of the complexes; see colour bar for values. The black dashed curve is the result of a third order polynomial fitted to the source density and thus traces the mean source latitude; This is provided to emphasise the shape of the Galactic warp.} 

\end{center}
\end{figure*}

We include the H$_2$ gas surface density derived from CO observations (\citealt{nakanishi2006}) on Fig.\,\ref{fig:rms_rgc_distribution} for comparison.  The molecular gas surface density peaks at $\sim$5\,kpc before decreasing exponentially with a scale length of $\sim$2.5\,kpc (as determined by \citet{nakanishi2006}). It is interesting to note that the source densities for the parts of the spiral arms inside the Solar circle are very similar (i.e., the peaks are approximately the same height $\sim$5-6\,kpc$^{-2}$), implying that the star formation rate per unit area is fairly constant. The source distribution is broadly similar to that of the molecular gas, with the peaks in the source distribution identifying localised regions where the SFE and/or SFR have been enhanced. The massive SFR surface density is therefore correlated with the molecular gas surface density. This is consistent with the SFR derived by \citet{misiriotis2006} from an analysis of COBE/DIRBE (1.2, 2.2, 60, 100, 140, 240\,\mum) and COBE/FIRAS (100-1000\,\mum) maps of the Galactic disk. Observations of nearby spiral galaxies (SINGS; \citealt{leroy2008}) found them to have a similar structure to that of the Milky Way with a approximately flat SFE towards their interiors, which decreased exponentially at large radii as the \hi/H$_2$ ratio increases. They infer that the SFRs do not appear to be directly sensitive to environmental conditions within giant molecular clouds (GMCs), but that the formation of GMCs is dependent on local conditions.  This would suggest that the spiral structures do not directly influence the SFR but are localized regions where the formation of GMCs is significantly more efficient.

While the correlation between the SFR and molecular gas surface densities is good over the whole range of available Galactocentric distances, the peak in the SFR density at $\sim$10\,kpc, associated with the fourth quadrant section of the \Sag\ arm, may be of particular note. The SFR surface density here is perhaps only half that of the segments of the spiral arms located within the Solar circle, but is projected against a much lower gas surface density, which may indicate that this spiral arm has a significantly enhanced SFR per unit gas mass beyond what is observed in the other spiral arms.  

The final feature to note is the minimum seen in the source density plots at approximately 9\,kpc. This minimum in the source counts changes by one bin in the distribution of the northern and southern samples and so is somewhat washed out in the distribution of the whole sample presented in the upper left panel of Fig.\,\ref{fig:rms_rgc_distribution}. This minimum roughly coincides with a similar feature identified by \citet{lepine2011} that they associated with a step down in metallicity and attributed to a kinematic barrier at co-rotation though which the gas cannot easily pass. Given the difficulties in assigning accurate distances to objects near the Solar circle, which have LSR velocities near zero, it would be speculative to draw conclusions regarding this feature at this stage, so we simply note the coincidence.

\subsubsection{Bolometric luminosity}

There are six peaks present in the luminosity distribution, four of which coincide with the peaks seen in the source count distribution (i.e., $\sim$3, 5, 6 and 10\,kpc).  Although the SFR surface densities are similar for the segments of the spiral arms located within the Solar circle, we find the luminosity surface densities increase with increasing distance from the Galactic centre.  As already mentioned, the overall SFR is similar for the for most of the arms within the Solar circle, so this increase in luminosity per unit area is likely to be due to an increase in the mean luminosity of the embedded stars. This would imply that the SFE is somehow enhanced in the outer parts of the spiral arms or that it is suppressed in the inner regions, perhaps by the interaction of the gas with the Galactic bar. 

We estimate the total luminosity of RMS embedded MYSO and \hii\ region populations by multiplying the luminosity in each bin by the area of the bin annulus and summing all of these together to obtain a value of $0.76\times10^8$\,\lsun. In this estimate we include all RMS sources, including those below the completeness in order to obtain a value that can be compared to that determined by \citet{bronfman2000} from their IRAS selected sample. Our value  is approximately half the value determined by \citet{bronfman2000}; however, their fluxes were drawn from the $IRAS$ point source catalogue, whose large beam sizes (i.e., $\sim$5\arcmin\ at 100\,\mum) would have likely caused an overestimate of the total luminosity. It is also possible that their sample will include a significant number of more evolved \hii\ regions. The luminosity of the embedded massive star population represents only a few per\,cent of the total Galactic far-infrared luminosity (i.e., $\sim2\times10^9$\,\lsun; \citealt[][$COBE$ DIRBE]{sodroski1997}; \citealt[][$IRAS$]{bloemen1990}), which would point to a relatively short embedded lifetime.

We use the total Galactic luminosity of the embedded massive star population to estimate the fraction contributed from each of the massive complexes discussed in Sect.\,\ref{sect:complexes}. This value is given in the final column of Table\,\ref{tbl:complex_parameters}. The full version of this table is only available in electronic form, however, the portion presented in this table lists the 25 most luminous complexes in the Galaxy. It is interesting to note that the 10 most luminous complexes contain only 8\,per\,cent of the embedded MYSOs and \hii\ region population, but contribute approximately 30\,per\,cent of the total luminosity of the Galactic embedded massive star population. A recent study of the WMAP free-free foreground emission maps  \citep{murray2010} found that the 18 most luminous Galactic star forming regions are responsible for the production of over half of the total ionising luminosity of Galaxy. This would suggest that the formation or early O-type stars is concentrated into a relatively small number regions in the Galaxy. Furthermore, this suggests that the initial mass function (IMF) is not constant on all scales and supports the cluster-based IMF model (e.g., \citealt{kroupa2003}).

Some of these complexes are clearly very luminous and contribute significantly to some of the peaks seen in Fig.\,\ref{fig:rms_rgc_distribution}. For example, the increased luminosity surface density seen in the northern distribution at 6\,kpc is somewhat due to the presence of the W51 star forming complex in this bin, which is the most active region in the Galaxy ($\sim$7\,per\,cent of the total RMS luminosity; see Table\,\ref{tbl:complex_parameters}). The additional peaks found at $\sim$7, 8 and 9\,kpc can be attributed to the G305, W49 and NGC\,3603 star forming complexes, respectively; three more of the most active in the Galaxy.

\subsubsection{\Sag\ arm}

We note that the \Sag\ arm is a prominent feature in all of the plots presented in Fig.\,\ref{fig:rms_rgc_distribution}, much more strongly detected and clearly defined in both hemispheres than, e.g. the Perseus arm.  The \Sag\ arm also shows up in the scale-height distribution presented in the lower panel of Fig.\,\ref{fig:z_rgc_distribution}. The peak seen at approximately 6.5\,kpc is only a feature of the northern Galactic plane sample and from the middle panels presented in Fig.\,\ref{fig:rms_rgc_distribution} the bin corresponding to this Galactocentric distance 
is coincident with the \Sag\ arm. This spiral arm therefore has a larger scaleheight than the other segments of the spiral arms located within the Solar circle.  

This prominence suggests that, as traced by massive star formation, the \Sag\ arm is a major feature of the Galaxy, rather than the minor arm implied by the sketch in Fig.\,6. This is contrary to the \citet{benjamin2008} conclusion, although the tracers are quite different as the GLIMPSE data traces an old stellar population so the implication may be that the SF history of the arms is periodic or variable.

\subsection{Galactic mid-plane and scaleheight distributions}

In Fig.\,\ref{fig:l_z_distribution} we show the latitude distribution of the whole RMS sample with respect to the Galactic mid-plane ($b=0$), as a function of Galactic longitude and radial velocity. This plot reveals that sources occupy a narrow range of distances ($|z|<100$\,pc) within the inner part of the Galactic plane (i.e., 300\degr $<\ell<$ 60\degr) and flare to significantly more positive and negative distances (up to 600\,pc in either direction) in the second and third quadrants, respectively. These deviations from the mid-plane follow the structure of the outer Galaxy, which is known to be significantly warped (\citealt{oort1958}). This is not the first example of infrared sources tracing the Galactic warp, this was previously reported by \citet[][see their Fig.\,3]{wouterloot1990}, but provides a nice confirmation of their results. 

The velocities of the sources associated with these large excursions from the mid-plane are reasonably coherent and tend towards the more extreme ends of the velocity distributions. Comparison with spiral-arm velocities (cf. Fig.\,\ref{fig:lv_distribution}) illustrates a correspondence with the Outer arm in the second quadrant and the Perseus arm in the third quadrant. The Perseus arm is also seen in the second quadrant, but lies close to the Galactic mid-plane with perhaps a small excursion to negative distances between $\ell=90$\degr\ and 120\degr. This corresponds to the part of the plane where the largest excursion to positive distances is seen towards the Outer arm.  The observed lower source density associated with the third quadrant part of the Perseus arm compared to the Outer arm is likely due to the larger heliocentric distances in this part of the Galaxy. 

There are also regions where the velocities are correlated within the inner Galaxy, and these can be matched to the expected spiral-arm tangents: Scutum-Centaurus, Sagittarius, Perseus, Norma, and Scutum-Centaurus at $\ell=$ 20\degr, 30\degr, 340\degr, 330\degr, 300\degr, respectively. We also note that nearly all of the complexes identified are located towards the Galactic mid-plane.

\begin{figure}
\begin{center}

\includegraphics[width=0.45\textwidth, trim= 0 0 0 0]{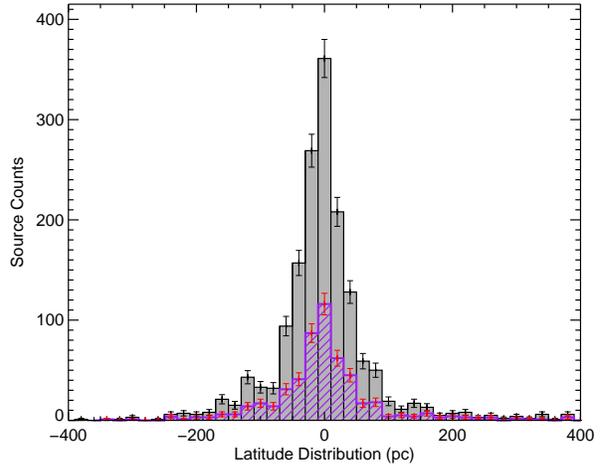}

\caption{\label{fig:z_distribution} Galactic latitude distribution of the embedded RMS population. The grey histogram shows the distribution of the whole sample while the purple hatched histogram show the subsample of sources above the completeness limit. The bin size is 20\,pc and the errors are determined from \poi\ statistics.} 

\end{center}
\end{figure}

In Fig.\,\ref{fig:z_distribution} we show the latitude distribution of the whole sample as well as the subsample of sources with luminosities above $2\times10^4$\,\lsun; scale heights obtained from fits to both samples agree within the errors with 37.7$\pm$0.8\,pc and 39.5$\pm$1.7\,pc, respectively. This suggests that the star-formation scale height is relatively luminosity independent over the range of luminosities covered by the RMS survey (i.e., $\sim$10$^2$-10$^6$\,\lsun). We also find the scale height is similar for the northern and southern Galactic-plane subsamples.

We note that these scale heights are significantly larger than those derived from the ATLASGAL-CORNISH sample of \uchii\ regions and their star-forming clumps ($\sim$22.4$\pm$1.9 and 28.1$\pm$2.6\,pc, respectively; Urquhart et al. 2013), or methanol masers (27$\pm$1\,pc; \citealt{green2011b}) identified by the MMB survey (\citealt{caswell2010c}) or indeed the subsample of RMS sources located within the GRS region (i.e., 30.2$\pm$2.8\,pc; \citealt{urquhart2011a}). However, most of these surveys have focused on sources located primarily within the inner Galaxy and so the larger scale height obtained for the $full$ RMS sample is simply be a consequence of including the outer Galaxy sources, which have a significantly larger distribution around the mid-plane (e.g., see Fig.\,\ref{fig:l_z_distribution}).

\begin{figure}
\begin{center}

\includegraphics[width=0.45\textwidth, trim= 0 0 0 0]{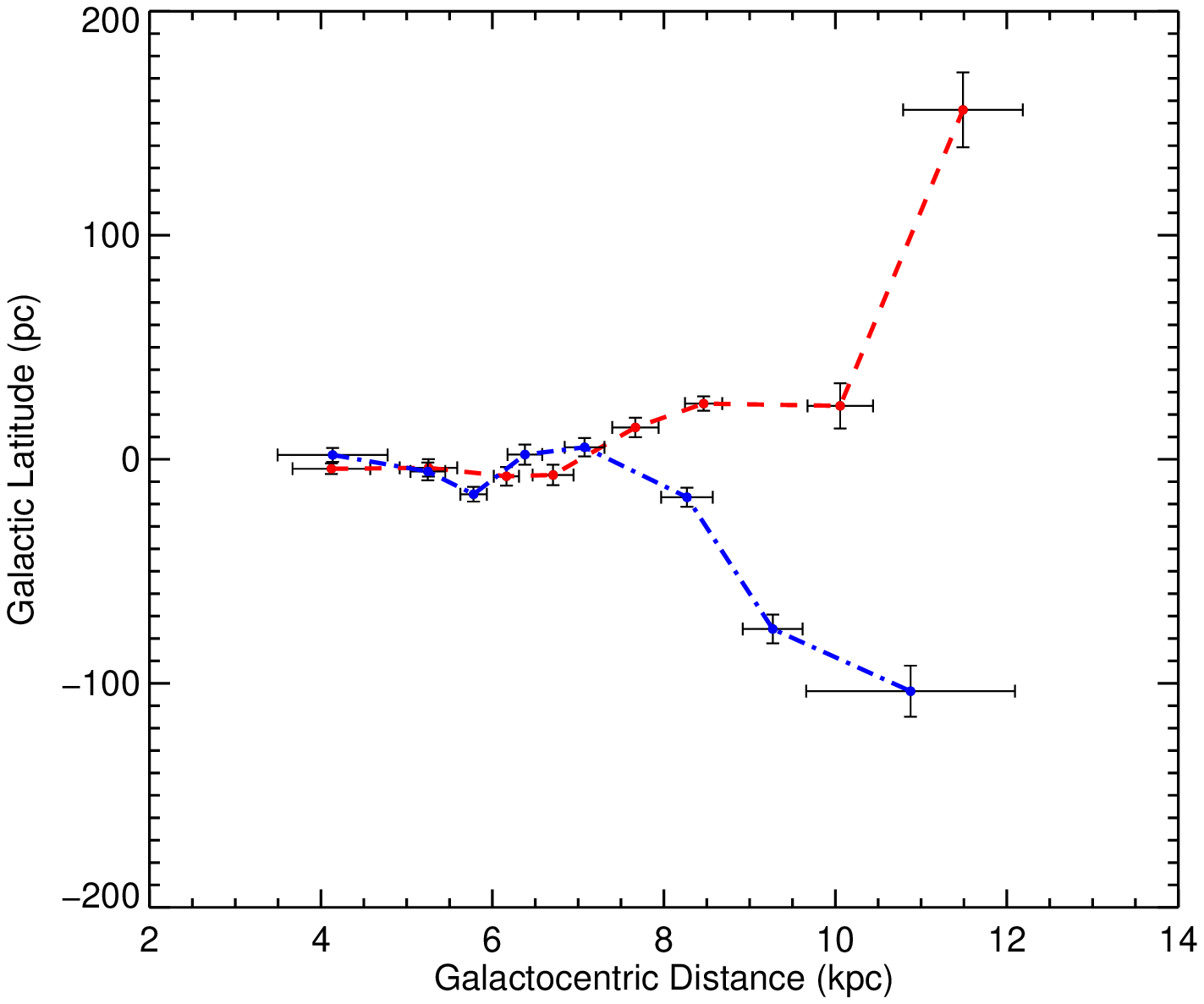}
\includegraphics[width=0.45\textwidth, trim= 0 0 0 0]{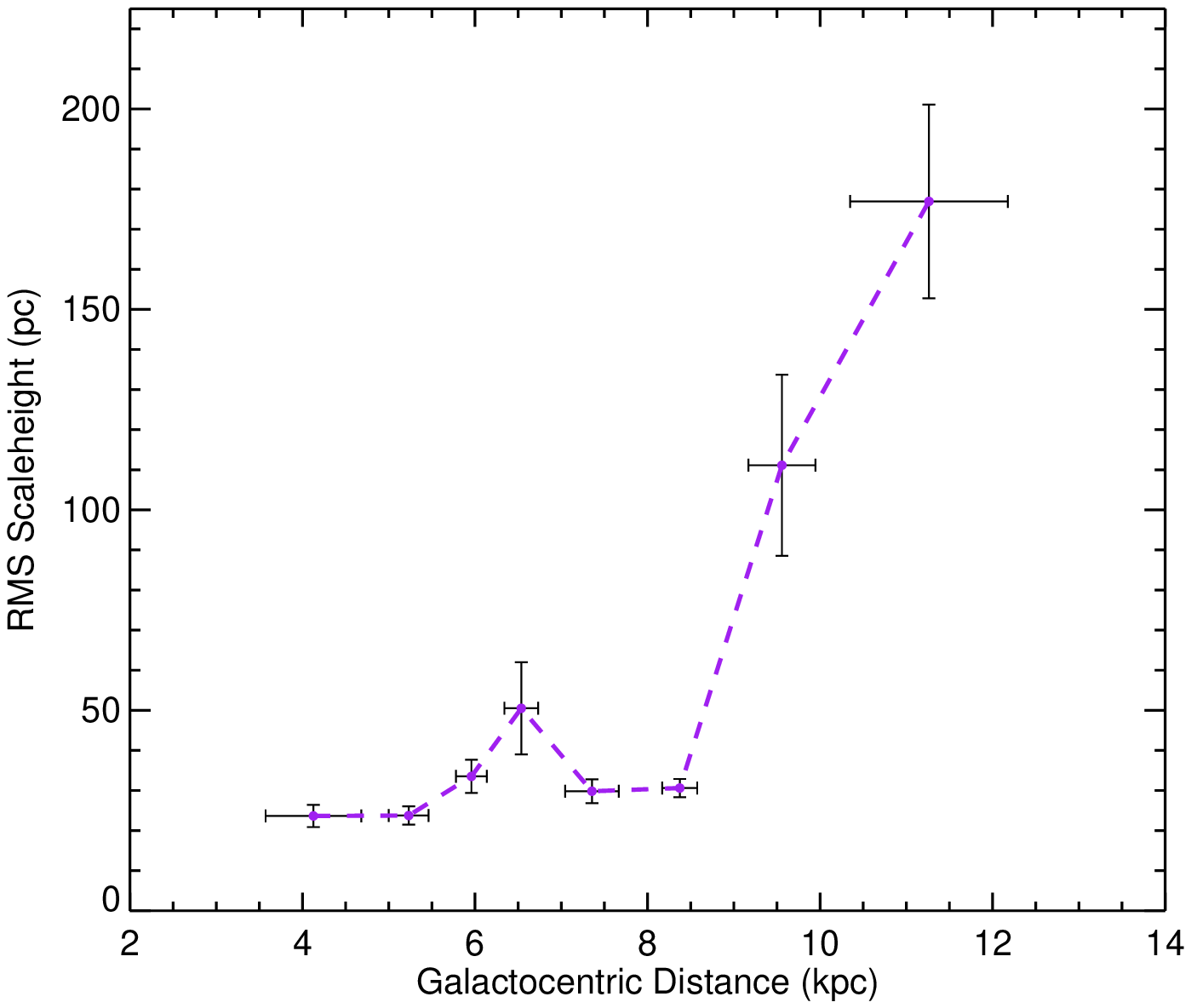}

\caption{\label{fig:z_rgc_distribution}  The offset of the mean RMS source position from the mid-plane and the scale height as functions of Galactocentric radius are shown in the upper and lower panels, respectively. The northern and southern Galactic plane subsamples are shown in red and blue in the upper panel, while in the lower panel we show only the distribution of the whole sample in purple.} 

\end{center}
\end{figure}

In Fig.\,\ref{fig:z_rgc_distribution} we present the offset of the Galactic disk from the mid-plane and measurements of the scale height as a function of Galactocentric radius.  The upper panel shows the distribution of the mean southern and northern subsample positions about the mid-plane separately to illustrate the effects of the Galactic warp.  The lower panel shows the absolute mean height from the mid-plane for the entire sample: the deflection of the southern plane below the mid-plane is similar to that of the northern plane above it. The samples for each of these plots have been divided into eight bins of equal number of sources. The distance from the mid-plane is determined from the average over each bin, and the error is the standard error on the mean. The scale height and its associated error are determined from an exponential fit to the histogram of data in each bin. The Galactocentric radius is simply the mean radius of the binned data, and the error is the standard deviation of this value.

These plots show that both the northern and southern samples are tightly associated with the Galactic mid-plane for radii $<$7\,kpc, after which they begin to diverge.
The Galactic scale height is also seen to increase modestly with increasing Galactocentric distance with a value of $\sim$20\,pc at a radius of 4\,kpc to $\sim$30\,pc at 8\,kpc.  At larger distances this slope increases very rapidly, reaching values of almost 200\,pc at a Galactocentric distance of $\sim$11\,kpc. We find no significant difference between the northern or southern samples. The increase in scale height with increasing Galactocentric radius is very closely matched by height of the molecular H$_2$ layer both in magnitude and slope (e.g., \citealt{nakanishi2006,wouterloot1990,bronfman1988, grabelsky1987}). Similar increases in scale height have been reported from \hi\ studies of the Milky Way (\citealt{malhotra1995}) and nearby edge-on spiral galaxies (\citealt{rupen1991}), although in these cases the \hi\ disks are significantly thicker, and increase much more quickly (100 to 220\,pc between 3 and 8\,kpc; \citealt{malhotra1995}).

\citet{lin1964} predict that the influence of spiral-arm shocks is significantly weaker in the outer Galaxy beyond the corotation radius ($R_{\rm{GC}}\sim 8$\,kpc). This would lead to less confined and smooth spiral features outside this radius, with supernovae probably playing a more important role than the spiral structure in determining the state of the ISM \citep{dibs2009}. Our RMS sources are young enough that they are still embedded in their natal molecular clouds and have not been dispersed by the Galactic gravitational potential, and are thus a good probe of changes in the molecular ISM. The relatively sharp change in the distribution and increase of the scaleheight of RMS sources outside the co-rotation radius would therefore support the idea of the ISM being generally more flocculent, dominated by supernovae and less confined in the outer Galaxy.

\setlength{\tabcolsep}{6pt}

\begin{table}

\begin{center}
\caption{Changes in scale height over the disk. }
\label{tbl:z_rgc}
\begin{minipage}{\linewidth}
\begin{tabular}{.....}
\hline
\hline

\multicolumn{1}{c}{$R_{\rm{GC}}$}	&	\multicolumn{1}{c}{\zo}&\multicolumn{1}{c}{$\Delta$\zo} &\multicolumn{1}{c}{Scaleheight} &\multicolumn{1}{c}{$\Delta$Scaleheight} \\  
\multicolumn{1}{c}{(kpc)}	&	\multicolumn{1}{c}{(pc)}&\multicolumn{1}{c}{(pc)} &\multicolumn{1}{c}{(pc)}&\multicolumn{1}{c}{(pc)} \\  

\hline
4.15 &-1.14&2.04 &22.32&1.49\\
5.27 &-7.26&2.63 &23.03&2.00\\
5.99 &-11.43&2.97 &31.57&3.35\\
6.55 &-1.01&3.07 &51.01&12.07\\
7.37 &6.37&3.44 &31.93&3.00\\
8.40 &3.53&2.85 &31.21&2.99\\
9.59 &-53.40&7.30 &113.89&23.79\\
11.29 &55.01&12.71 &176.82&24.82\\

\hline
\end{tabular}
\end{minipage}

\end{center}

\end{table}
\setlength{\tabcolsep}{6pt}

\section{Summary and conclusions}
\label{sect:summary_conclusions}

The Red MSX Source (RMS) survey is a Galaxy-wide mid-infrared selected sample of $\sim$1750 MYSOs and \uchii\ regions. In this paper we summarise the results of our previous molecular line follow-up observations, and present additional observations towards $\sim${\color{black}800} RMS sources. Observations of the $^{13}$CO (1-0) transition were made towards sources not previously observed, and the higher density tracers CS and NH$_3$ were used to identify the correct radial velocity for sources towards which multiple velocity components had been detected. 

Combining these results with those from our previous follow-up campaign of observations and with archival line surveys, we have obtained radial velocities to all but a handful of sources. Use of the positional and radial velocities of our sources combined with inspection of mid-infrared images has allowed us to group sources into $\sim120$ star forming complexes. Approximately one third of the RMS sample are associated with these complexes.

We have conducted an in-depth literature search to identify sources/complexes for which a reliable distance has previously been determined. Where a distance was not available we have used the radial velocities and the Galactic rotation curve derived by \citet{reid2009} to calculate kinematic distances, and have used the \hi\ self-absorption  to resolve any outstanding kinematic distance ambiguities for sources within the Solar circle ($\sim$200 sources). We have obtained distances to $\sim$1650 \hii\ regions and YSOs, which corresponds to over than 90\,per\,cent of the embedded RMS population. These distances are used to estimate bolometric luminosities using model fits to the spectral energy distributions previously derived by \citet{mottram2011a}. We find we are complete above $2\times10^4$\,\lsun\ to a heliocentric distance of $\sim$18\,kpc.

We use this sample to investigate the Galactic distribution of massive stars with respect to the position of the spiral arms, the Galactic long bar, the Galactic warp and the flaring of the molecular disk. Finally we compare the distribution of massive stars in our Galaxy with other nearby spiral galaxies. Our main findings are as follows:

\begin{enumerate}

 \item We observe a high surface density of MYSO and \hii\ regions within the Solar circle and a decaying exponential at larger distances from the Galactic centre, consistent other measurements of the star formation rate determined from infrared and submillimetre studies.  Variation in the star formation rate and efficiency within the Solar circle appears to be attributable to specific large scale feature of Galactic structure, such as the spiral arms and the ends of the bar.

 \item We  have compared the distribution of the RMS sources with the expected position of the spiral arms using longitude-velocity diagrams, in 3-dimensions and as a function of Galactocentric radii and have found them to be in good agreement. Our results are therefore consistent with a model of the Galaxy consisting of four major arms.
 
 \item The embedded source and H$_2$ gas surface densities have a very similar overall distribution. We find that the source surface densities associated with the segments of the spiral arms located within the Solar circle are very similar, which suggests that the massive star formation rate per unit molecular gas mass is approximately constant for the inner parts of the spiral arms. 
  
 \item We also find that the luminosity surface density increases with increasing Galactocentric radius for the segments of the spiral arms located within the Solar circle. Given that the source surface densities are approximately the same for the spiral arms in this region, this indicates that the mean source luminosity is increasing with distance from the Galactic centre. This increase in mean source luminosity can be attributed to a very small number of extremely active star forming complexes (i.e., W51, W49, W43) where the star and/or clump formation efficiency is significantly enhanced.
 
 \item We find the scaleheight of massive star formation as measured from the whole sample is 39.5$\pm$1.7\,pc, which is larger than has been reported previously ($\sim$30\,pc).  Most earlier studies have concentrated on inner Galaxy samples and exclude the significant flaring of the disk observed in the outer Galaxy. We measure the scaleheight as a function of Galactocentric distance and find that it increases only modestly between $\sim$4 and 8\,kpc (i.e., from $\sim$20-30\,pc), but much more rapidly at larger distances. 
 
 \item  We estimate the total integrated bolometric luminosity of the embedded MYSO and \hii\ region population to be $\sim0.76\times 10^8$\,\lsun; however, we find that the ten most luminous complexes contribute almost 30\,per\,cent of the total integrated RMS bolometric luminosity while comprising only 8\,per\,cent of the sources.
 
\end{enumerate}

\section*{Acknowledgments}
 
We would like to thank the referee for their informative comments and suggestions that have improved this work. We would also like to extend thanks to Friedrich Wyrowski for reading and commenting on an earlier draft of the manuscript. The Mopra radio telescope is part of the Australia Telescope National Facility which is funded by the Commonwealth of Australia for operation as a National Facility managed by CSIRO. The University of New South Wales Digital Filter Bank used for the observations with the Mopra Telescope was provided with support from the Australian Research Council. We thank Dr. Mark Reid for providing the Fortran code used to estimation of kinematic distances for their Galactic rotation curve. This research has made use of the SIMBAD database operated at CDS, Strasbourg, France and NASA's Astrophysics Data System Bibliographic Services. This work was partially funded and carried out within the Collaborative Research Council 956, sub-project A6, funded by the Deutsche Forschungsgemeinschaft (DFG). This paper made use of information from the Red MSX Source survey database at {\tt{http://rms.leeds.ac.uk/cgi-bin/public/RMS\_DATABASE.cgi}} which was constructed with support from the Science and Technology Facilities Council of the UK.

\bibliography{cornish}

\bibliographystyle{mn2e_new}

\end{document}